\documentclass[12pt]{article}     

\usepackage{geometry}
\usepackage{graphicx}
\usepackage{amsmath}
\usepackage{lscape}
\usepackage{multirow}
\usepackage{float}

\usepackage{array}

\usepackage{subcaption} 

\usepackage{multicol}

\usepackage[numbers]{natbib}
\usepackage{booktabs}
\usepackage[inline]{enumitem}

\usepackage{color}
\usepackage{textcomp}
\usepackage{url}

\usepackage[dvipsnames]{xcolor}

\usepackage{pifont}
\usepackage{soul}
\soulregister\cite7
\soulregister\ref7
\soulregister\pageref7

\date{}

\title{\Large \textbf{Diurnal temperature variation as the source of the preferential direction of fractures on asteroids: theoretical model for the case of Bennu}}

\author{D. Uribe$^{a, b,}$\thanks{\ Corresponding author. \newline  \indent \indent E-mail address: diego.uribe\_suarez$@$mines-paristech.fr (D. Uribe Su\'{a}rez).} , M. Delbo$^b$, P.-O. Bouchard$^a$ ,  D. Pino Mu\~{n}oz$^a$}

\begin{document}           

\maketitle
\vspace{-1.2cm}
\begin{center}
$^a$ Centre de Mise en Forme des Mat\'eriaux (CEMEF)\\
MINES ParisTech, PSL Research University\\
CNRS UMR 7635, CS 10207 rue Claude Daunesse, 06904 Sophia Antipolis Cedex, France\\
$^b$ Universit\'e C\^ote d'Azur, Observatoire de la C\^ote d'Azur\\
CNRS-Lagrange, CS 34229, 06304 - NICE Cedex 4, France
\end{center}  
	

\begin{center}
\section*{Abstract}
\end{center}
It has been shown that temperature cycles on airless bodies of our Solar System can cause damaging of surface materials. Nevertheless, propagation mechanisms in the case of space objects are still poorly understood. Present work combines a thermoelasticity model together with linear elastic fracture mechanics theory to predict fracture propagation in the presence of thermal gradients generated by diurnal temperature cycling and under conditions similar to those existing on the asteroid Bennu. 
The crack direction is computed using the maximal strain energy release rate criterion, which is implemented using finite elements and the so-called $G\theta$ method (Uribe-Su{\'a}rez et al. 2020. Eng. Fracture Mech. 227:106918). 
Using the implemented methodology, crack propagation direction for an initial crack tip in different positions and for different orientations is computed. It is found that cracks preferentially propagate in the North to South (N-S), in the North-East to South-West (NE-SW) and in the North-West to South-East (NW-SE) directions. Finally, thermal fatigue analysis was performed in order to estimate the crack growth rate. Computed value is in good agreement with available experimental evidence.\\ 

\noindent
\textit{Keywords:} Asteroid Bennu, Thermoelastic model, Crack propagation Direction, Energy Release Rate, Thermal fatigue crack growth 

\section{Introduction}
The formation of fractures and their growth are key processes in man-made structures and materials. The large majority of the cases studied in the literature are on the aforementioned structures, given their importance in the everyday life of humans. These processes also occur in natural objects, such as  rocks, boulders, and cliffs \citep{Cao2019, AlMukhtar2015, Meng2019, Atkinson1982, Vastola2011, Eppes2015NatCo...6E6712E} and are documented on several objects of our solar system, including Earth \citep{Collins2016NatGe...9..395C,Collins2018NatCo...9..762C}, Mars \citep{Eppes2015NatCo...6E6712E,Viles2010GeoRL..3718201V}, our Moon \citep{Ruesch2020Icar..33613431R,Li2017P&SS..145...71L}, the nuclei of comets \citep{Attree2018A&A...610A..76A,Matonti2019NatGe..12..157M, Maarry2015}, asteroids \citep{Dombard_2010,Lauretta2019Natur.568...55L,Walsh2019NatGe..12..242W,DellaGiustina2019NatAs...3..341D,Molaro2020NatCo..11.2913M}, and meteorites \citep{Delbo2014Natur.508..233D}.

The source of the driving forces that provokes crack nucleation and propagation can be very diverse, ranging from unloading of the pressure stresses under which certain rocks formed in the deep crust of Earth, tectonic stresses, rapid mechanical stresses from impacts and thermal stresses. In the latter case, the presence of water can enhance the cracking phenomena via the known freeze-thaw effect \citep[see e.g.][and references therein]{Hall2004ESPL...29...43H}. The effectiveness of thermal stresses in cracking rocks and other geological units in the absence of water has been long debated: a famous laboratory experiment by \citet{Griggs1936JG.....44..783G} argued against earlier claims that rocks could be fractured by temperature variations only, the process that in general and hereafter is called thermal cracking. However, thermal cracking gained momentum recently and came to great attention to planetary scientists thanks to new measurements, modelling, and observations. 

More in details, it has been shown that temperature variations resulting from the cycles between day and night can damage materials on airless bodies of our Solar System \citep[for references on this topic, see the introductions of ][]{Molaro2017Icar..294..247M,Molaro2015JGRE..120..255M}. This damaging process consists in the nucleation and growth of micro-fractures inside the material due to the mechanical stresses induced by the diurnal temperature cycles. In general the mechanical stresses resulting from temperature gradients due to the day and night cycles are smaller than the strength of the material \citep[see e.g.][]{Delbo2014Natur.508..233D,Ravaji2019JGRE..124.3304R,ElMir2019Icar..333..356E,Hazeli2018Icar..304..172H,Molaro2017Icar..294..247M}. In this case the crack can still open and grow in a regime that is said to be sub-critical \citep{Atkinson1984JGR....89.4077A}; the material is increasingly damaged at each cycle and it is usually spoken of thermal fatigue. Eventually, the application of a large number of cycles can produce important crack growth; the crack tip can reach a boundary of the material, such as a discontinuity or the edge of a rock; this will cause a rapid transition from sub-critical to critical failure 
and lead to material failure \citep[see e.g.][for the case of asteroids and meteorites] {Dombard2010Icar..210..713D, Delbo2014Natur.508..233D}. For example, \citet{Liang_2020} studied volumetric stress distribution in an L6 ordinary chondrite's microstructure subjected to thermal and mechanical loadings through the combination of experiments and micromechanical models. It was found that under thermal cycling, the stress concentrates more uniformly along with particle interfaces. The authors interpret that thermal fatigue crack propagation could result in the debonding of particles from the surrounding matrix.

Hence, on solar system bodies without an atmosphere thermal fatigue of surface rocks, in addition to the impact of micrometeorites, can eventually lead to rocks' breakup and produce regolith \citep{Dombard2010Icar..210..713D, Delbo2014Natur.508..233D}, the latter being the layer of unconsolidated material that covers planetary surfaces \citep{Yano2006, Veverka2001, Murdoch2015aste.book..767M}. In the case of the near-Earth asteroid (101955) Bennu \citep{Lauretta2019Natur.568...55L}, \citet{Molaro2020JGRE..12506325M} propose that thermal cracking is also able to eject sub-cm-sized particle away from the asteroid surface, thereby offering an explanation for the observed activity of this asteroid \citep{Lauretta2019Sci...366.3544L}. Furthermore, it is also proposed by several studies that macroscopic fractures, mass-wasting, and material breakdown on asteroids and cometary nuclei could be explained as a consequence of thermal effects \citep{Dombard2010Icar..210..713D, Maarry2015, Ali2015, Attree2018A&A...610A..76A,Molaro2020NatCo..11.2913M}.

For all the reasons above, thermal fatigue cracking is now considered a space weathering mechanism. On the other hand, direct evidence of thermal cracking on asteroids (and comets) is still relatively scarce (but strongly growing) and the details of the process in terms of spatial and temporal scales still poorly understood. One of the first studies that invokes this phenomenon to explain certain \emph{in situ} asteroid observations is the work of \cite{Dombard2010Icar..210..713D}.Using images obtained by NASA's Shoemaker mission of the surface of the asteroid (433) Eros, these authors noted boulders that appear to break and erode in place, producing fragments that fill the inside of craters, creating characteristic ``ponds'' of regolith. Another observational evidence, obtained from images of NASA's OSIRIS-REx mission \citep{Lauretta2014M&PS..tmp..113L},  is constituted by the detection of exfoliation sheets on some of the boulders on the asteroid (101955) Bennu \citep{Molaro2020NatCo..11.2913M}. The thickness of the exfoliation sheets is consistent with the depth inside boulders at which thermoelastic simulations show stress concentration as a result of diurnal  temperature variations \citep{Molaro2020NatCo..11.2913M}. However, it has been shown on Earth \citep{McFadden2005GSAB..117..161M} and Mars \citep{Eppes2015NatCo...6E6712E} that  one of the most diagnostic observations of thermal cracking induced by diurnal temperature variations is a preferential meridional direction (north to south) of the fractures on surface rocks.  The reason is very simple: during the day the Sun moves in the sky from the east to the west. As a consequence, the temperature gradients are directed essentially in the same direction (west to east). Fractures mainly propagate in a direction perpendicular to that of maximum principal stress. Therefore it is expected that this direction of propagation is essentially from the north to the south, when they are driven by these diurnal temperature cycles. There are hints of predominance of fracture directed in the north to the south and in the north-west to the south-east on the boulders of the asteroid Bennu \citep{Delbo2019EPSC...13..176D}. However, a modelling of the crack propagation direction in conditions similar to those existing on Bennu is still lacking. The aim of this work is to provide theoretical foundation for analysis and interpretation of fracture directions on small asteroids with properties similar to those of Bennu. For our simulations, a thermo-mechanical model will be used to predict  stresses on Bennu's boulders and an appropriate crack propagation method based on principles of fracture mechanics.

The paper is organised as follows. In Section 2, the developed thermomechanical model is presented. This section also reviews crack growth direction theory using an energetic approach for linear thermoelasticity fracture mechanics problems. One simple example is presented in Section 3 to show the accuracy of the proposed method when calculating the crack propagation direction due to the presence of thermal gradients. The results are compared against observed crack propagation directions on asteroid Bennu. In section 4 the computed crack propagation directions are discussed. Finally some concluding remarks are presented in Section 5.

\section{Methods}
\label{methodssection}

Primary goal of this work is to compute the direction of crack propagation due to thermal strain in a geometry corresponding to a typical boulder on the surface of the asteroid (101955) Bennu, the target of NASA's sample return mission OSIRIS-REx. Here it is made the hypothesis that most of the fractures observed on the surface of the boulders by \cite{Lauretta2019Natur.568...55L}, \cite{DellaGiustina2019NatAs...3..341D}, and \cite{Delbo2019EPSC...13..176D} are due to the growth of surface cracks. \cite{Molaro2020JGRE..12506325M} also discuss stresses inside boulders that could cause  cracking deep inside the rock mass. 

\begin{figure}[h!]
	\centering
	\includegraphics[width=\textwidth]{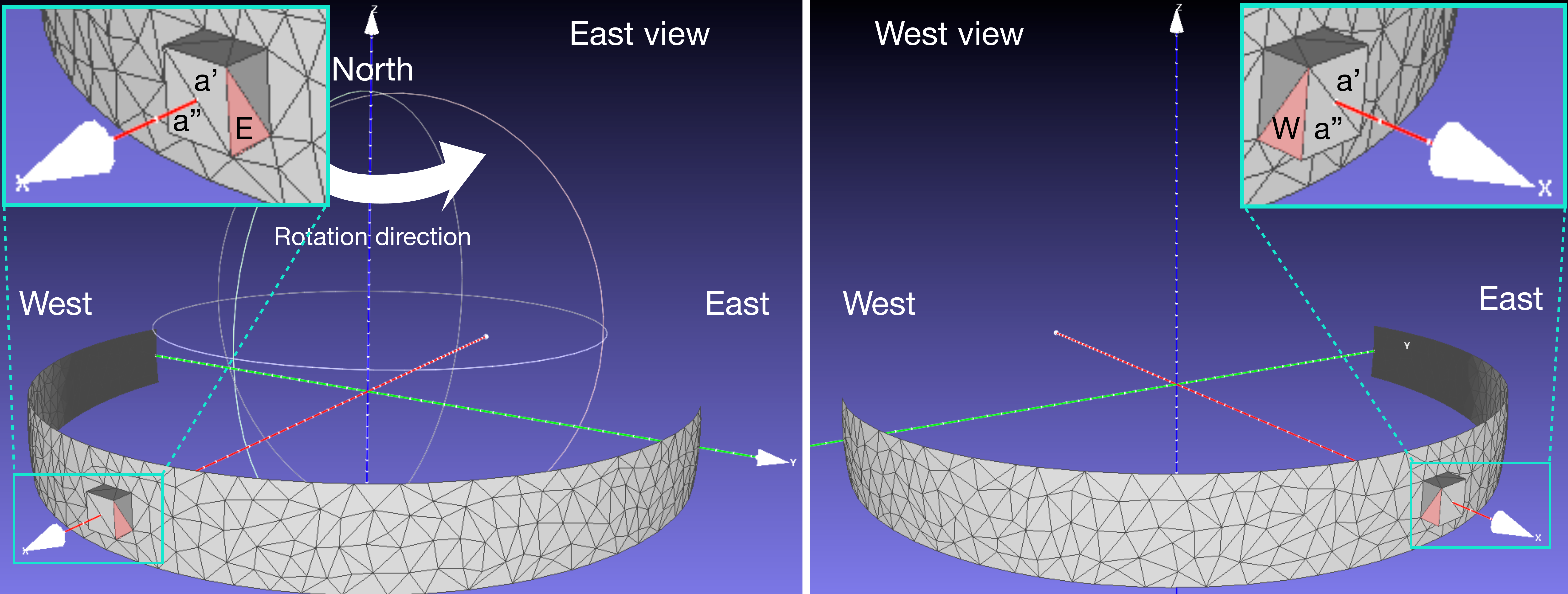}
	\vspace{-0.1cm}
	\caption{Schematic representation of the cubic-like boulder extruding from the equator of the asteroid, of which only about half of the equatorial belt is simulated to obtain the temperature distribution on faces \textbf{E} and \textbf{W}.}
	\label{F:geometry}
\end{figure}

The geometry of our problem is schematised in Fig.~\ref{F:geometry}, where it is shown a cubic-like boulder extruding from the equator of the asteroid, of which only about half of the equatorial belt is simulated. 

The mesh is divided in triangular facets. A thermophysical model \citep{Delbo2015aste.book..107D} is used to calculate the temperatures of all facets as a function of time, as described in the following sections. The temperatures of the boulder facets are then used in a thermoelastic model in order to compute the strain and stresses as a function of the position in the boulder and time. Next, fracture mechanics theory is used to estimate the propagation direction of a tiny notch that is placed on the horizontal (a', a") face of the cubic boulder. This is done by computing the strain energy release rate \footnote{The strain energy release rate represents the change of elastic strain energy per unit area of crack extension.}, which is a well-known fracture mechanics parameter \citep{griffith_1921}. The implemented thermoelastic model as well as the used methodology for the computation of the energy release rate ($G$) are explained in the following sections.

\subsection{Thermo-physical model}
The first step of the presented method consists in using a well-established thermophysical model \citep{Spencer1989Icar...78..337S,Delbo2015aste.book..107D} to solve the one-dimensional heat diffusion problem. Temperature is calculated as a function of time for all the surface elements of the mesh depicted in Fig.~\ref{F:geometry}. Boundary conditions are given by the variable day/night illumination including the shadows cast by the local terrain of the mesh on itself, radiation of the heat in space, conduction in the sub surface, and mutual heating \citep{Rozitis2011MNRAS.415.2042R}.
The physical parameters of the material used in this work are given in Table~\ref{T:param}. These properties were taken from \cite{Delbo2014Natur.508..233D}. They correspond to the Carbonaceous Chondrite sub-type CM2 Murchison meteorite. This is considered to be a good analog of asteroids belonging to the C-complex broad spectroscopic class \citep{DeMeo2009Icar..202..160D}.  The asteroid Bennu also belong to the C-complex \citep{Hamilton2019NatAs...3..332H}.

\smallskip

\begin{table}[h!]
\centering
\begin{tabular}{cccc}
\toprule
\toprule
      Quantity name, symbol   & Units               & Value & Reference           \\
\midrule  
    Rotational period, $P$       &  s   &  15,469.2  &   [1] \\
    Bulk modulus, $K$       &  MPa   &  29,000  & [2]  \\
    Shear modulus, $\mu$  & MPa           & 18,000  & [2]    \\
    Young modulus, $E$  & MPa &   44,742.857  & [2]        \\
    Poisson's ratio, $\nu$  &  & 0.2428  & [2]   \\
    Bulk density, $\rho$  & kg \ m$^{-3}$ & 1,662  & [2]   \\
    Thermal conductivity, $\lambda$  & W \ mm$^{-1}$ \ K$^{-1}$ & 5 $\times$ 10$^{-4}$  & [2]    \\
    Thermal expansion coefficient, $\alpha$  & K$^{-1}$ & 8.5 $\times$ 10$^{-6}$  & [2]  \\
    Heat capacity, $c$  & J \ kg$^{-1}$ \ K$^{-1}$ & 500  & [2]   \\
Reference temperature, $T_{ref}$  & K & 250  & [3]   \\
Paris pre-factor, $C$  & m  [MPa $\sqrt{\text{m}}$]$^{-n}$ & 3 $\times$ 10$^{-4}$   & [2]  \\
Paris exponent, $n$  &  & 3.84  & [2]  \\
\bottomrule
\bottomrule
\end{tabular}
\caption{Thermal and mechanical properties and their default values used in this work for the simulations of the crack propagation directions on asteroid Bennu. References: [1] = \cite{Barnouin2019NatGe..12..247B}, [2] = \cite{Delbo2014Natur.508..233D}, [3] = this work.}
\label{T:param}
\end{table}

\subsubsection{Thermoelastic model}

\begin{figure}[h!]
\centering
\includegraphics[width=1.0 \textwidth]{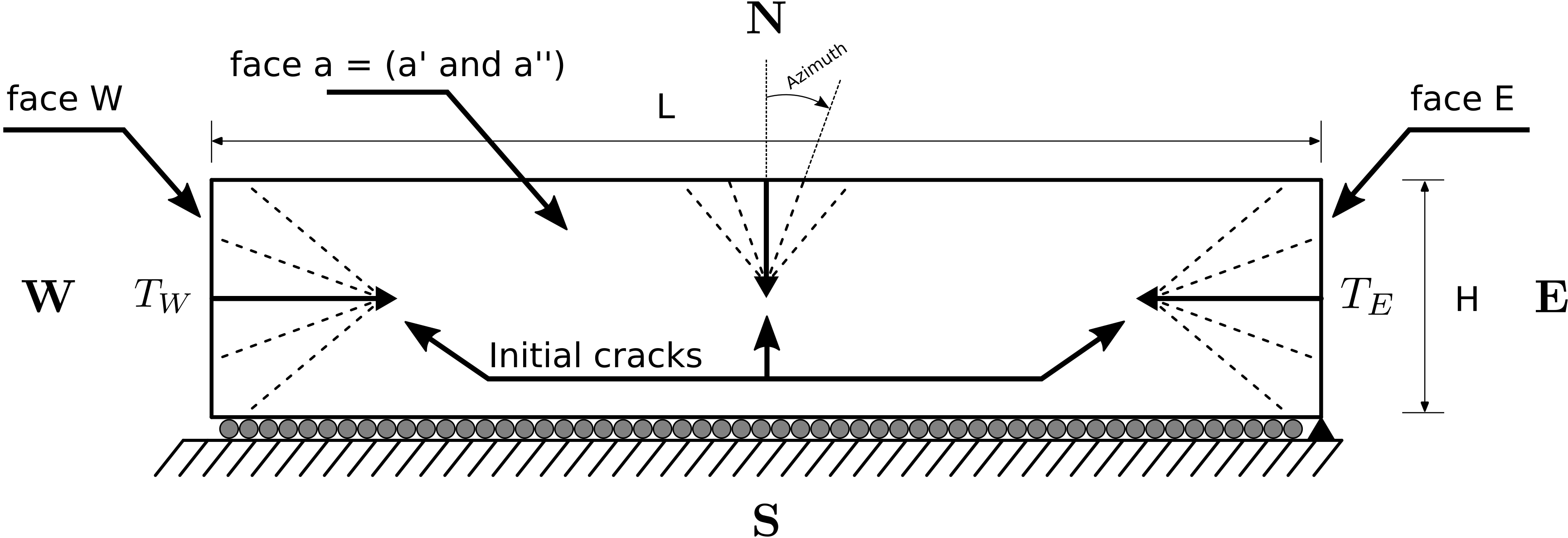}
\caption{Geometry and boundary conditions for the face of the boulder parallel to the surface. Crack tip position and crack axis were varied. The length of the a-face is also varied in our simulations.}
\label{GeometryBeamAsteroidBennu}
\end{figure}

\citet{Delbo2019EPSC...13..176D} observed and mapped cracks on boulders on the surface of the asteroid Bennu using OSIRIS-REx images that can be approximated as parallel to the local surface of the asteroid. On the other hand \cite{Delbo2014Natur.508..233D}, \cite{Ravaji2019JGRE..124.3304R} and \cite{ElMir2019Icar..333..356E} considered cracks propagating perpendicularly to the local surface.
For this reason, current work is interested in the component of the crack growth parallel to the local  surface of the asteroid, namely the plane perpendicular to the X-axis of the mesh of Fig.~\ref{F:geometry}. To study this case, a 2-D model where the initial crack is placed on the a' and a" facets of Fig.~\ref{F:geometry} is used. These two facets are treated together in the following as the single planar face that is called ``a-face'' (Fig.~\ref{GeometryBeamAsteroidBennu}). The separation of the a-face in two facets is required to make the implemented thermo-phyisical model, that uses triangular facets only, compatible with the meshing algorithm.

The vertical E and W facets of the 3D mesh of the thermo-physical model (Fig.~\ref{F:geometry}) are respectively  mapped to the vertical right and left sides of the beam simulated by the proposed thermoelastic and fracture mechanics model (Fig.~\ref{GeometryBeamAsteroidBennu}).

The crack propagation on the a-face is essentially driven by the 2D temperature gradient created by the strong temperature mismatch between the E- and W-face, it will be explained later on in Section \ref{ResultsSection}. These temperature gradients exist throughout the whole day/night cycles and strongly depend on time. There is also a component of the temperature gradient perpendicular to the a-face, which causes crack propagation in the sub-surface. This component has been studied by \cite{Delbo2014Natur.508..233D}, \cite{ElMir2019Icar..333..356E} and \cite{Ravaji2019JGRE..124.3304R} providing laws to estimate the crack growth rate in that direction.


Here, therefore, only the temperature gradient is considered in the a-face. To do so, the heat diffusion equation is used to calculate the temperature as a function of space and time in the a-face given as boundary conditions the temperatures on the E- and W-faces, which are determined by the asteroid thermo-physical model.

When temperature variation takes place inside an elastic continuum, it generally induces thermal stresses. This phenomenon can be taken into account through the linear addition of thermal strains to mechanical ones. 
Once the thermal problem has been solved, the effect of the temperature variation on the mechanical response of the body can be included in the total strain ($\epsilon$), which can be decomposed into an elastic, plastic and thermal part, as it is shown in equation (\ref{totalstrain}):

\begin{equation}
\label{totalstrain}
\epsilon_{ij} \ = \  \epsilon_{ij}^{elastic} \ + \epsilon_{ij}^{plastic} \ + \ \epsilon_{ij}^{thermal};
\end{equation}

\noindent
where $\epsilon_{ij}$ is the total strain tensor, $\epsilon_{ij}^{elastic}$, $\epsilon_{ij}^{plastic}$ and $\epsilon_{ij}^{thermal}$ are respectively, the elastic, plastic and thermal strain tensors. In this work it is assumed that there is no plastic strain ($\ \epsilon_{ij}^{plastic} = 0$). Therefore, equation (\ref{totalstrain}) simplifies into:

\begin{equation}
\label{totalstrain_noplastic}
\epsilon_{ij} \ = \  \epsilon_{ij}^{elastic} \ + \ \epsilon_{ij}^{thermal};
\end{equation}

\noindent
where thermal strain is defined as:

\begin{equation}
\label{thermalstrain}
\epsilon_{ij}^{thermal} \ = \ \alpha \Delta T \delta_{ij} \ , \ \ \  \ where \ \Delta T = T - T_{ref}
\end{equation}

\noindent
In equation (\ref{thermalstrain}), $\alpha$ is the thermal expansion coefficient $[K^{-1}]$, $\Delta T$ is the difference between the computed temperature and the reference temperature, latter one being the temperature where there is no strain. Finally $\delta_{ij}$ is the Kronecker delta ($\delta_{ij}$ = 0 for i $\neq$ j, $\delta_{ij}$ = 1 for i = j). It is assumed that the reference temperature is the average of the temperatures given in Section \ref{ResultsSection}. It is also assumed an isotropic thermal expansion coefficient.


In the presented method a weak thermomechanical coupling is assumed. Which means that the temperature is initially obtained from the heat problem and then introduced into the mechanics computation. This is possible since the characteristic time scale of the thermal problem is several orders of magnitude greater than the characteristic time scale of the crack propagation problem. For the solution of a thermoelastic problem, first, the temperature distribution inside the body is computed by solving the heat transfer problem. Then, the resulting temperature distribution is input to the mechanical problem as an initial strain. In elasticity, the Hooke's law for homogeneous and isotropic materials is defined by Equation (\ref{elasticityequation}):

\begin{equation}
\label{elasticityequation}
\sigma_{ij} = \lambda \delta_{ij} \epsilon_{kk} + \mu (\epsilon_{ij} + \epsilon_{ji}) 
\end{equation}

\noindent
where $\sigma_{ij}$ is the stress tensor and $\lambda$ and $\mu$ are respectively, the Lam{\'e}'s first and second parameters. When thermal strain is included, Equation (\ref{elasticityequation}) becomes:

\begin{equation}
\label{thermoelasticityequation}
\sigma_{ij} = \lambda \delta_{ij} (\epsilon_{kk} - 3 \alpha \Delta T) + 2 \mu (\epsilon_{ij} - \alpha \Delta T \delta_{ij}) 
\end{equation}

\subsection{Energy Release Rate (G)}

Crack propagation requires energy. The amount of energy released during the fracture process is known as energy release rate ($G$). Commonty, $G$ is widely used in the literature in order to find the crack propagation direction for a given configuration. To compute the crack propagation direction, a criterion based on the azimuthal distribution of the energy release rate around the crack tip is used. Multiple crack propagation directions are tested and then select the one that maximises $G$. Namely, in the plane of the a-face, the azimuth of the maximum energy release rate is identified, as it is known that crack propagation will take place in that direction. This method was tested and validated in \citet{Uribe_2020}.

The $G$-value is evaluated using all virtual and kinematically admissible crack length displacements. The direction of crack propagation ($\theta_{0}$) can be determined by:

\begin{equation}
\label{DerivativeCriteria}
\begin{cases}
\left(\frac{d G}{d \theta}\right)_{\theta = \theta_{0}} = 0, \\
\left(\frac{d^2 G}{d \theta^2}\right)_{\theta = \theta_{0}} \leq 0. 
\end{cases}
\end{equation}              

\noindent
In equation (\ref{DerivativeCriteria}), according to the work done by \citet{erdogan_crack_1963}, there is a limit angle corresponding to pure shear: $\theta = \pm 70.54^{\circ}$. In this work, for the computation of the energy release rate ($G$), the numerical technique known as $G\theta$ method \citep{Destuynder1983} is used. The $G\theta$ method's implementation is quite simple and multiple extensions are available.

The strain energy release rate is the decrease in the total potential energy ($w_{p}$) during a growth of crack area ($dA$). To determine the variation of the total potential energy in cracked solid $\Omega$, $F^{\varepsilon}$ is defined as an infinitesimal geometrical perturbation $\varepsilon$ in the vicinity of the crack tip:

\begin{equation}
\label{Perturbation}
\begin{aligned}
& F^{\varepsilon} : \mathbf{R}^{3} \rightarrow \mathbf{R}^{3} \\
& \forall M \in \Omega, \ F^{\varepsilon}(M) = M^{\varepsilon} = M + \varepsilon \textbf{V} (M)
\end{aligned}
\end{equation}

\noindent
where the virtual field $\textbf{V}$ gives the location of each point of the perturbated solid using its initial position ($M$) before the perturbation. When the perturbation $\varepsilon$ is sufficiently small, \citet{Destuynder1983} showed that the stress field ($\sigma$) and the displacement field ($u$) on the perturbed configuration may be expressed as:

\begin{equation}
\label{Perturbated_fields}
\begin{aligned}
& \sigma^{\varepsilon} = \sigma + \varepsilon \sigma^{1} \\
& u^{\varepsilon} = u + \varepsilon u^{1}
\end{aligned}
\end{equation} 

\noindent
where $\sigma^{1}$ and $u^{1}$ are the first order variations of the stress and displacement fields during the infinitesimal perturbation $\varepsilon$ on $\Omega$. The total potential energy variation during a crack extension is then obtained when $\varepsilon$ aims towards 0:

\begin{equation}
\label{totalPotenener}
\frac{d W_{p}}{da} = \lim_{\varepsilon \to 0} \frac{W^{\varepsilon}_{p}-W_{p}}{\varepsilon}
\end{equation}

The virtual displacement field $\textbf{V}$ representing the virtual kinematics of the crack has the following properties:

\begin{itemize}
  \item $\textbf{V}$ is parallel to the crack plan.
  \item $\textbf{V}$ is normal to the crack front.
  \item The support of $\textbf{V}$ is only needed in the vicinity of the crack.
  \item $\left\| \textbf{V} \right\|$ is constant in a defined region around the crack tip.
\end{itemize}

\begin{figure}[h!]
	\centering
	\includegraphics[width=0.83 \textwidth]{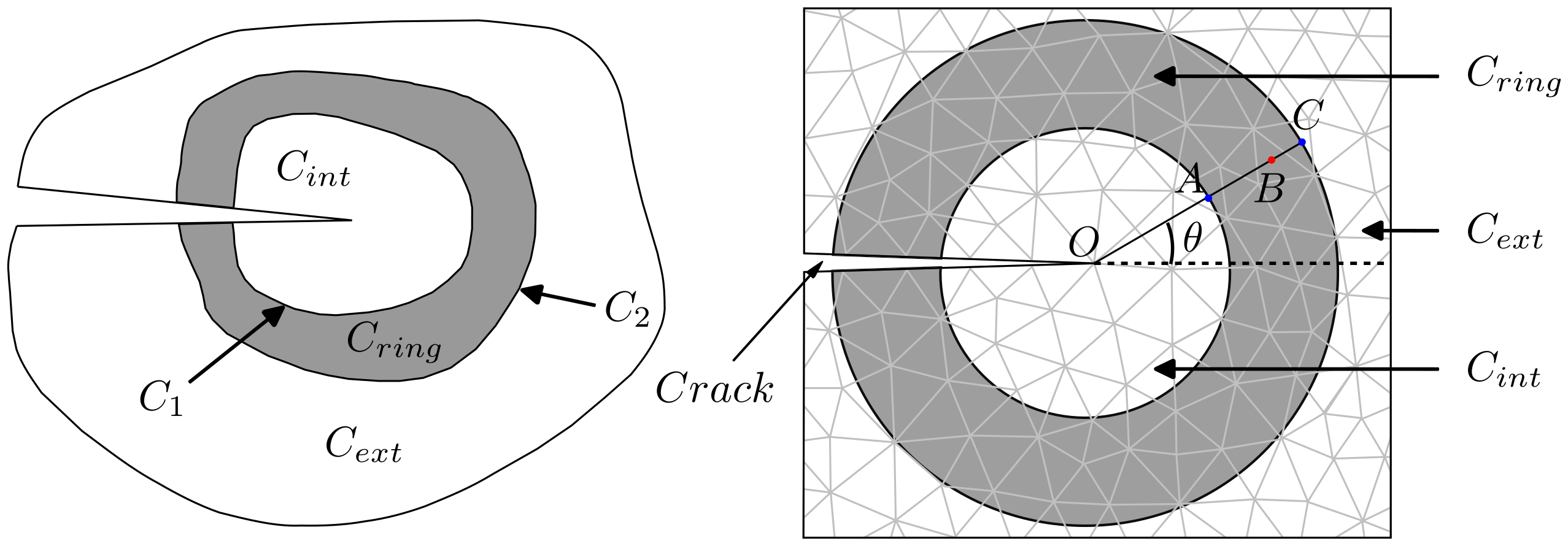}
	\vspace{-0.1cm}
	\caption{Contours and domains used to computed $G$ using the $G\theta$ method.}
	\label{contour_GTheta}
\end{figure}

When implementing the $G\theta$ method, first, it should be defined two contours $C_{1}$ and $C_{2}$ around the crack tip. Those contours divide this region into three domains $C_{int}$, $C_{ring}$ and $C_{ext}$ as it is shown in Fig. \ref{contour_GTheta}.

The virtual displacement field, $\textbf{V}(v_{1},v_{2})$, is defined by equation (\ref{virtual_displa}).   

\begin{equation}\label{virtual_displa}
  \textbf{V} = \begin{cases} 
	v_{1} = \left(1 - \frac{AB}{AC}\right) \cos( \theta)  \\
	\\
	v_{2} = \left(1 - \frac{AB}{AC}\right) \sin( \theta)  \\
  \end{cases}
\end{equation}

\noindent
Where $O$ is the crack tip, $B$ is an integration point belonging to the ring, $A$ and $C$ are the intersections between $OB$ and $C_{1}$ and $C_{2}$ respectively (inside and outside contours of the ring); and $\theta$ is the virtual direction of propagation measured with respect to the crack axis. Propagation direction is defined as a function of an angle $\theta$. Hence the values of $\textbf{V}$ in the three domains are:

\begin{itemize}
  \item The norm of $\textbf{V}$ in $C_{int}$ is constant and equal to 1.
  \item The norm of $\textbf{V}$ in $C_{ring}$ varies continuously from 1 to 0.
  \item The norm of $\textbf{V}$ in $C_{ext}$ is equal to 0.
\end{itemize}

\begin{figure}[h!]
	\centering
	\includegraphics[width=0.88 \textwidth]{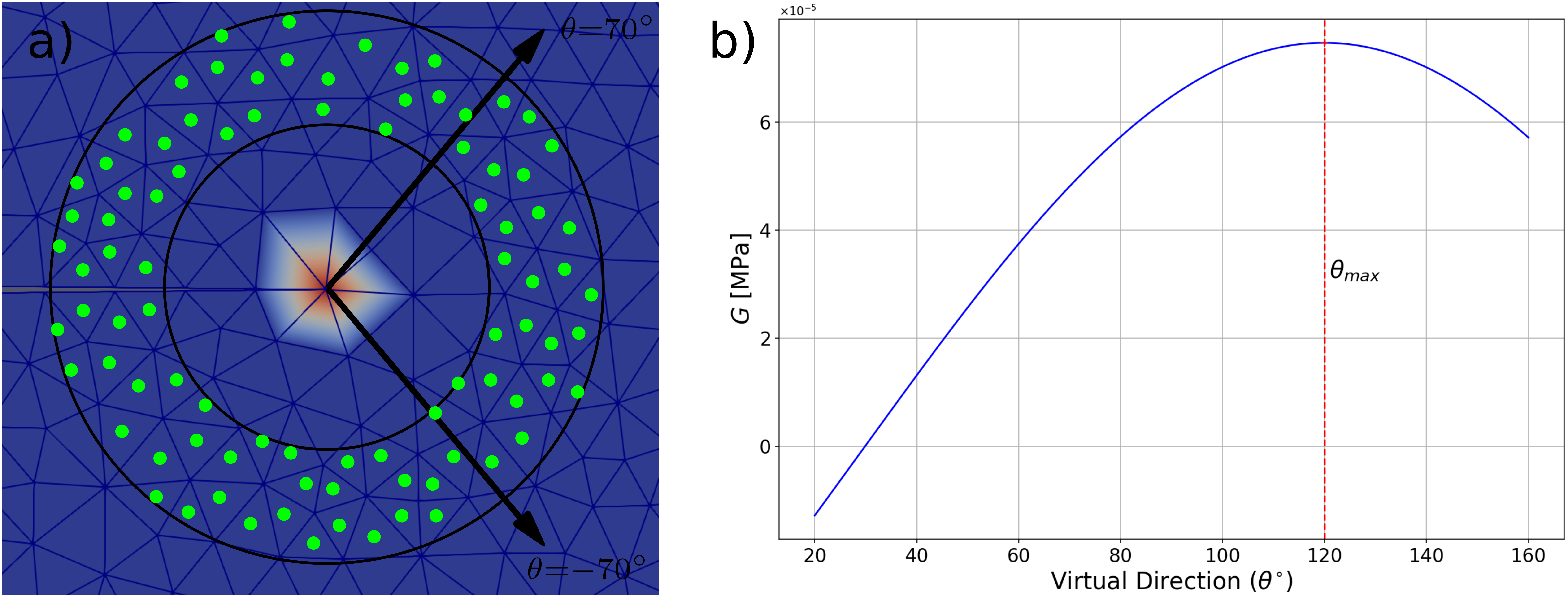}
	\caption{a) Ring of elements around the crack tip. b) $G(\theta)$ curve for the maximun energy release rate criterion for a crack where its crack axis equal to 90$^{\circ}$.}
	\label{ring_elements_gthetaCurve}
\end{figure}

When there is neither thermal strain nor load applied directly to the crack faces, the total potential energy may be expressed as:

\begin{equation}\label{totalpotentialenergyIntegral}
W_{p} = \frac{1}{2} \int_{\Omega} \sigma_{ij} U_{j,i} \ d \Omega - \int_{\Omega} f_{i} U_{i} \ d \Omega
\end{equation}

\noindent
where $\sigma_{ij}$ is the stress field, $U_{i}$ is the displacement field, 
and $f_{i}$ the external loads. When an infinitesimal perturbation $\varepsilon$ takes place, derivatives and integrals on the perturbed part can be expressed using a first order ``limited development'' of operations related to the non-perturbed part. Due to this, the energy release rate may be expressed as: 

\begin{equation}\label{energy_release_equ}
G =\int_{ring} \left( \sigma_{ij} U_{j,k} V_{k,i} - \frac{1}{2} \sigma_{ij} U_{j,i} V_{k,k} \right) dA_{ring}
\end{equation}

\noindent
where $\textbf{V}_{i}$ is the virtual displacement field, $U_{j,k}$ is the gradient of the displacement field, $V_{k,i}$ is the gradient of the virtual displacement field, $V_{k,k}$ is the divergence of the virtual displacement field and $A_{ring}$ is the integration region. Additionally, as $\textbf{V}_{i}$ varies only inside $C_{ring}$, the integration will only be performed over $C_{ring}$ ($A_{ring}$). In Fig. \ref{ring_elements_gthetaCurve}-\textbf{a}, elements belonging to $C_{ring}$ are marked with a green dot. A discrete set of $\theta$ values will be used for a virtual crack propagation. The $\theta$ values are selected inside the range $[-70^{\circ},70^{\circ}]$ \citep{erdogan_crack_1963} with respect to the crack axis. For each one of the $\theta$ values, an associated $G$ value will be computed. In Fig. \ref{ring_elements_gthetaCurve}-\textbf{a} the virtual ring in which all the integration points are considered is depicted. Fig. \ref{ring_elements_gthetaCurve}-\textbf{b} shows that $G$ can be computed for each value of $\theta$, which makes the identification of the $\theta$ value that maximises $G(\theta)$ straightforward. Fig. \ref{ring_elements_gthetaCurve}-\textbf{b} shows the curve $G(\theta)$ for a crack where its crack axis is equal to 90$^{\circ}$ (E), in this case the direction of propagation is equal to 120$^{\circ}$ (NW - SE).

Furthermore, due to the presence of thermal strain in this work, the inclusion of additional terms in the $G\theta$ method is necessary. In elasticity according to Hooke's law, the stress and strain tensors can be defined as follows:

\begin{equation}\label{stresstensordefinitionsstrain}
\sigma_{ij} = C_{ijkl} \ \epsilon_{kl}
\end{equation} 


\noindent
where $C_{ijkl}$ is the fourth-order elastic constitutive tensor. Combining equations (\ref{stresstensordefinitionsstrain}), (\ref{totalstrain_noplastic}) and (\ref{thermalstrain}) leads to the definition of the stress tensor taking into account thermal strain:

\begin{equation}\label{stresstensordefinitionthermalstrain}
\sigma_{ij} = C_{ijkl} \ \left( \epsilon_{kl} \ - \ \alpha \Delta T \delta_{kl} \right)
\end{equation} 

\begin{figure}[h!]
	\centering
	\includegraphics[width=0.535 \textwidth]{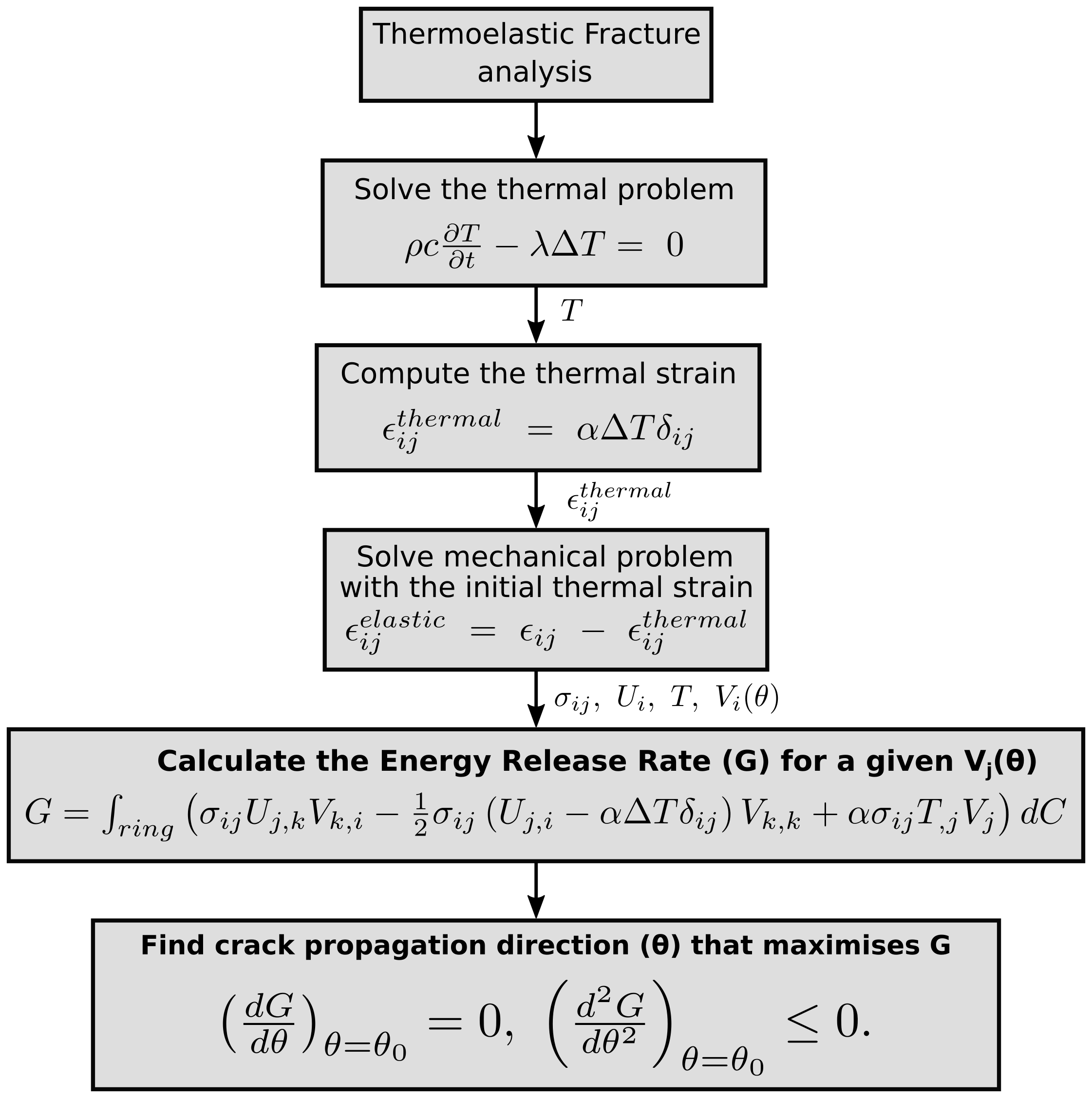} 
	\caption{Schematic representation for the thermoelastic fracture analysis performed in this work.}
	\label{schematicRepresentation}
\end{figure}

When thermal strain is accounted for, equation (\ref{Perturbated_fields}) is still valid. Therefore, according to \citet{Suo_1992,Brochard_1994}, the equation for the $G\theta$ method in the present of thermal strain is given:

\begin{equation}\label{Thermal_energy_release_equ}
G =\int_{ring} \left( \sigma_{ij} U_{j,k} V_{k,i} - \frac{1}{2} \sigma_{ij} \left(U_{j,i} - \alpha \Delta T \delta_{ij}\right) V_{k,k} + \alpha \sigma_{ii} T_{,j} V_{j} \right) dA_{ring}
\end{equation}

\noindent
where $\alpha$ is the thermal expansion coefficient, $\Delta T$ ($\Delta T = T - T_{ref}$) is the difference between the computed temperature and the reference temperature, $\delta_{ij}$ is the Kronecker delta and $T_{,j}$ is the gradient of the temperature. A schematic representation of the sequential solution of the thermomechanical problem presented in this work is provided in Fig. \ref{schematicRepresentation}.

\subsection{Fatigue crack growth model}

For linear elastic materials, $K$ and $G$ are uniquely related as follows \citep{anderson_2005}:

\begin{equation} \label{GlobalEquivalenceKG}
G = \frac{K_{I}^2}{E^{'}} + \frac{K_{II}^2}{E^{'}} + \frac{K_{III}^2}{2 \mu } \ \ \textrm{with} \ \ \begin{cases}
               E^{'} = E               & \ \textrm{for plane stress}\\
               E^{'} = \frac{E}{1 - \nu^{2}}     & \ \textrm{for plane strain}\\
               \ \mu = \frac{E}{2(1 + \nu)}     
           \end{cases} \\ 
\end{equation}

\noindent
where $G$ is the energy release rate, $K_{I}$, $K_{II}$ and $K_{III}$ are respectively the stress intensity factors for mode $I$, mode $II$ and mode $III$, $E$ is the Young modulus, $\nu$ is the Poisson's ratio and $\mu$ is the shear modulus. Due to the fact that present work deals with thermal cyclic stress (tension-compression), only mode $I$ loading is assumed. Once the energy release rate ($G$) is known, it is possible to compute the stress intensity factor ($K_{I}$). Assuming plane strain conditions, a relation between $G$ and $K_{I}$ can be stated from equation (\ref{GlobalEquivalenceKG}): 

\begin{equation}
K_{I} = \sqrt{\frac{G \ E}{1-\nu^2}}
\label{EquivalenceKG}
\end{equation}

\noindent
where $K_{I}$ is the stress intensity factor, $G$ is the energy release rate, $E$ is the Young modulus and $\nu$ is the Poisson's ratio. Paris's Law \citep{Janseen_2004} allows to relate the rate at which the crack length ($a$) grows as a function of the number of cycles ($N$) to the maximum variation of stress intensity factor ($\Delta K_{I}$) over the whole cycle. Paris's law is presented in equation (\ref{ParisLaw}):

\begin{equation}
\frac{da}{dN} \ = \ C [\Delta K_{I}]^{n}
\label{ParisLaw}
\end{equation}

\noindent
where $a$ is the crack length $[m]$, $N$ the number of cycles $[-]$, $\Delta K_{I}$ the range of stress intensity factor $[Pa \ m^{0.5}]$, $C$ $[m \ (Pa \ m^{0.5})^{-n}]$ and $n$ $[-]$ are material properties fitted to experimental fatigue data.

The beam presented in Fig. \ref{GeometryBeamAsteroidBennu}, also referred as the a-face in this work, is embedded in a mesh that is different from the one showed in Fig. \ref{F:geometry}. The aforementioned mesh is an isotropic unstructured triangular one, that is refined in the neighborhood of the crack tip (Fig. \ref{ring_elements_gthetaCurve}-\textbf{a}). The thermoelasticity problem is solved on this mesh through the finite element method (FEM). The developed thermoelastic model has been implemented in CimLib, a C\texttt{++} in-house finite element library developed at CEMEF {\citep{Cimlib2007}}. Regarding the thermal problem, the available finite element framework in CimLib is a classic one where the variable temperature ($T$) is solved through an implicit formulation \citep{Ryan2020}. The mechanical problem is solved using an implicit formulation (mixed) with first-order elements using linear interpolation functions for both velocity and pressure. In this formulation both variables are $P1$. This element combination, linear in both $\overrightarrow{v}$ and $p$ ($P1/P1$), does not satisfy the $inf-sup$ condition {\citep{Arnold1984, brezzi1991mixed}}. The latter, being a condition related with the mathematical convergence characteristics of the finite element formulation. The $inf-sup$ condition ensures the solvability, stability and optimality of the finite element solution and is crucial in establishing its quality \citep{Bathe_2001}. Through the $inf-sup$ condition, a problem can be assessed in order to see if it is well-posed or not. A problem is called well-posed if: \begin{enumerate*}[label=(\roman*)]
\item a solution exists,
\item the solution is unique and
\item the solution depends continuously on the given data (in some reasonable topology)\end{enumerate*} \citep{Hadamard_1902}. To overcome this issue, an extra degree of freedom (DOF) in the center of each element is added in $\overrightarrow{v}$. This extra DOF is condensed and thus written as a function of the remaining DOFs of the element. The use of this extra DOF is a well-known stabilization technique often called ``bubble'' or $P1^{+}/P1$. This technique satisfies the $inf-sup$ condition, and has been implemented in the CimLib {\citep{Rachid_2010,Cao2013}}.

\section{Results}
\label{ResultsSection}

\begin{figure}[h!]
\centering
\includegraphics[width=0.82 \textwidth]{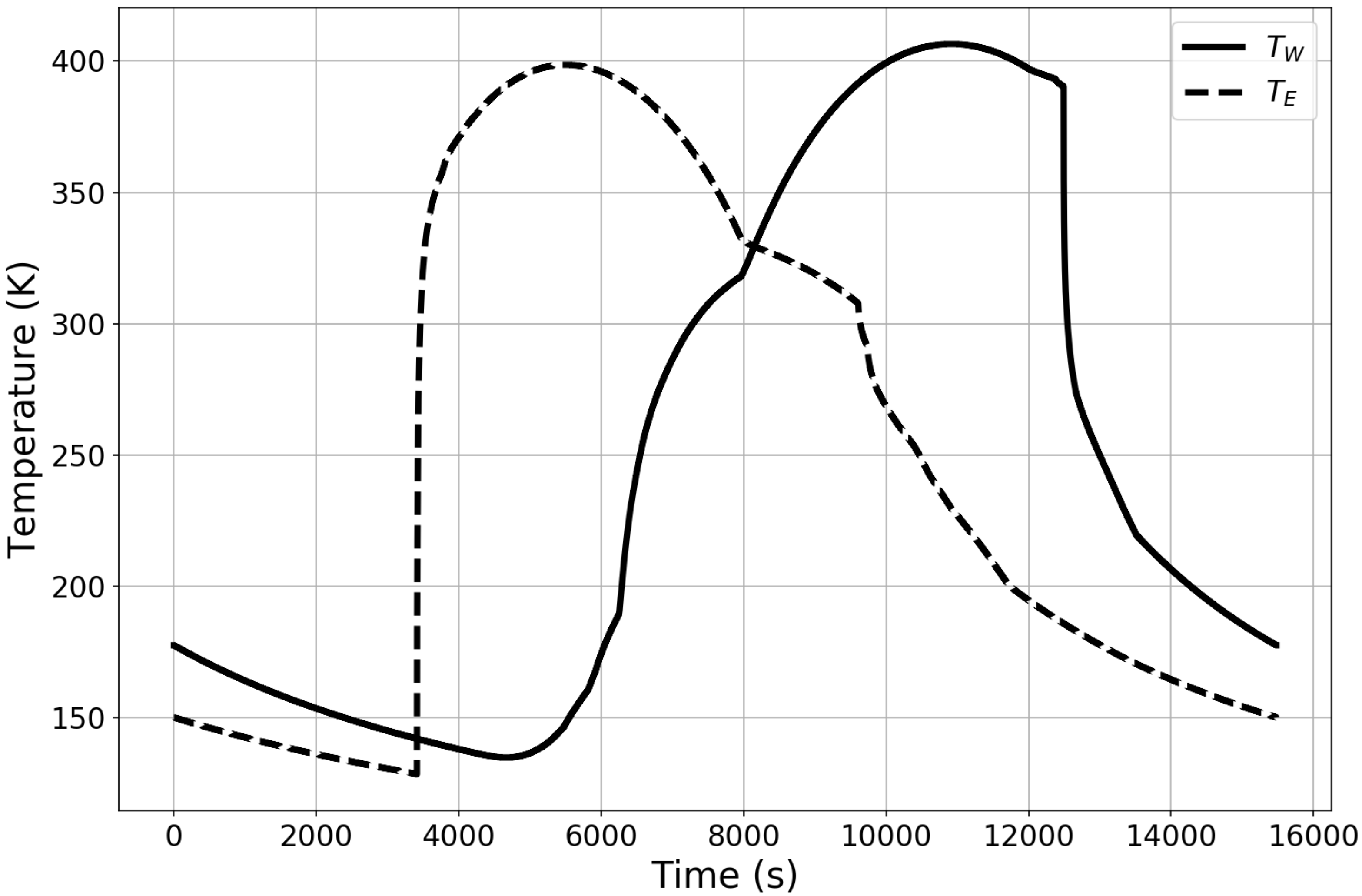}
\caption{Temperatures $T_{W}$ (W-face) and $T_{E}$ (E-face) used as thermal boundary conditions.}
\label{ThermalBC}
\end{figure}

In this work, the simulations begin by calculating temperatures for the mesh of Fig. \ref{F:geometry} using the asteroid thermophysical model described in section \ref{methodssection}. The asteroid is placed at a distance of 1.12 $au$ from the sun. This corresponds to the semi major axis of the orbit of the asteroid Bennu. The resulting temperatures of the E- and W-faces are shown in Fig. \ref{ThermalBC}. Next, these temperatures are input into the thermo-mechanical model. This second model predicts the stress over the domain of the a-face as a function of space and time. Since the presence of cracks highly affects the stress field, different simulations for different crack positions and orientations for the same temperature fields are run.

Namely, the propagation direction for an initial crack in different positions inside the domain of the a-face is computed, as shown in Fig. \ref{GeometryBeamAsteroidBennu}. In addition, for each position of each initial crack, its orientation is varied. Namely, when the crack is attached to the W-face the crack propagation direction is computed for different initial orientations with azimuth angles ranging from $46^{\circ}$ to $134^{\circ}$. Azimuth angle is increasing clockwise (i.e. from the north to the east) and is equal to zero when it is pointing vertical up (i.e. to the north). When the crack is attached to the E-face, the crack propagation direction is simulated for different initial crack orientations having azimuth angles ranging from $226^{\circ}$ to $314^{\circ}$. Finally, when the crack is attached to the north (top) side of the a-face (also called beam), the crack propagation direction is simulated for different initial crack orientations with azimuth angles ranging from $136^{\circ}$ to $224^{\circ}$. In all the described cases, the length and the width of the a-face were fixed to 100 $mm$ and 4 $mm$ respectively. These values were chosen after studying: \begin{enumerate*}[label=(\roman*)]
\item the characteristic time of the thermal problem as well as\label{item:CharaTime}
\item the minimum required width of the a-face in order to avoid the boundary effects on the results and \label{item:BoundaryEffect}
\item to avoid a high computational cost.\label{item:CompTime}\end{enumerate*} Due to the fact that there is no temperature gradient in the vertical direction in the aforementioned simulations, a larger height of the domain does not affect initial crack propagation direction. Similarly, this work does not intend to evolve the crack with time.

The results are presented using the so-called windrose diagrams, which can be considered as circular histograms that represent the distribution of the computed crack propagation directions. It should be noticed that for a rotation of $180^{\circ}$ of the circular histograms (windroses) these are identical. This is due to the fact that crack propagation direction (azimuthal angle) has to be independent of the beginning and ending point of a crack, i.e., for each one of the computed crack propagation directions $\theta$, in the histograms there are also count the direction $\theta + 180^{\circ}$. For the scenario where crack propagation direction is computed for an initial crack tip placed in different positions inside the domain of the a-face and for different crack orientations, the windroses are presented separately. 

\begin{figure}[h!]
\centering
\begin{subfigure}[b]{0.32\textwidth}
    \centering
    \includegraphics[width=\textwidth]{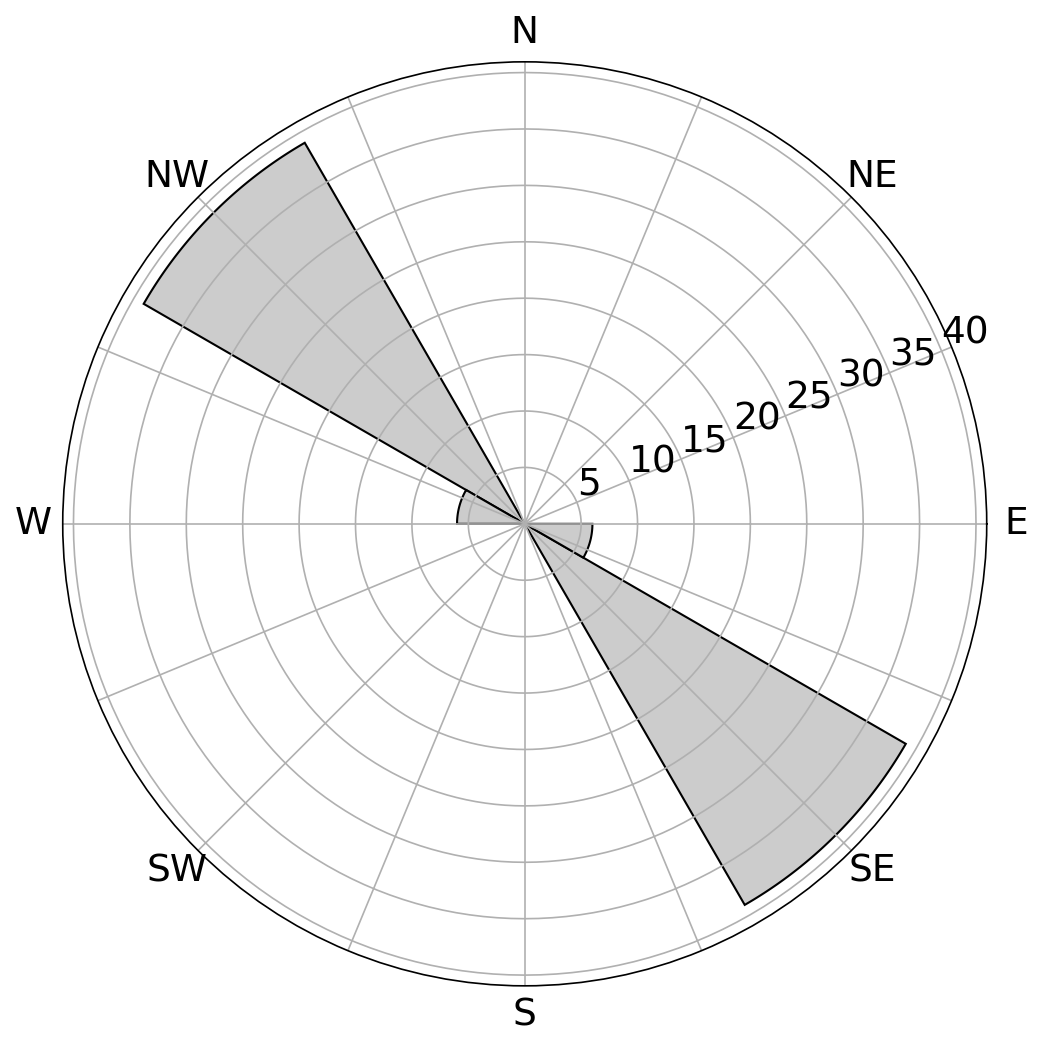}
    \caption{W-face}\label{CrackLeft}
\end{subfigure}
\begin{subfigure}[b]{0.32\textwidth}
    \centering
    \includegraphics[width=\textwidth]{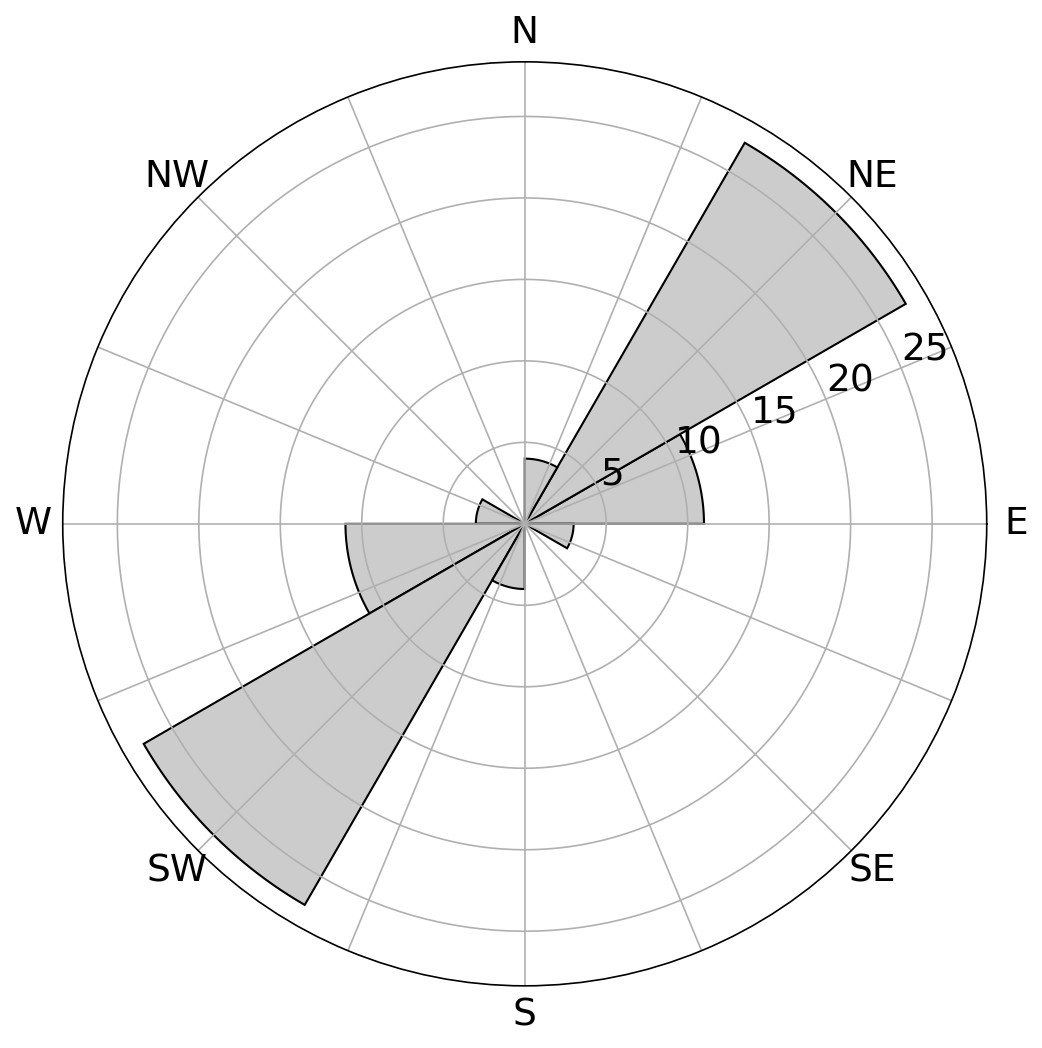}
    \caption{E-face}\label{Crackright}
\end{subfigure}
\begin{subfigure}[b]{0.32\textwidth}
    \centering
    \includegraphics[width=\textwidth]{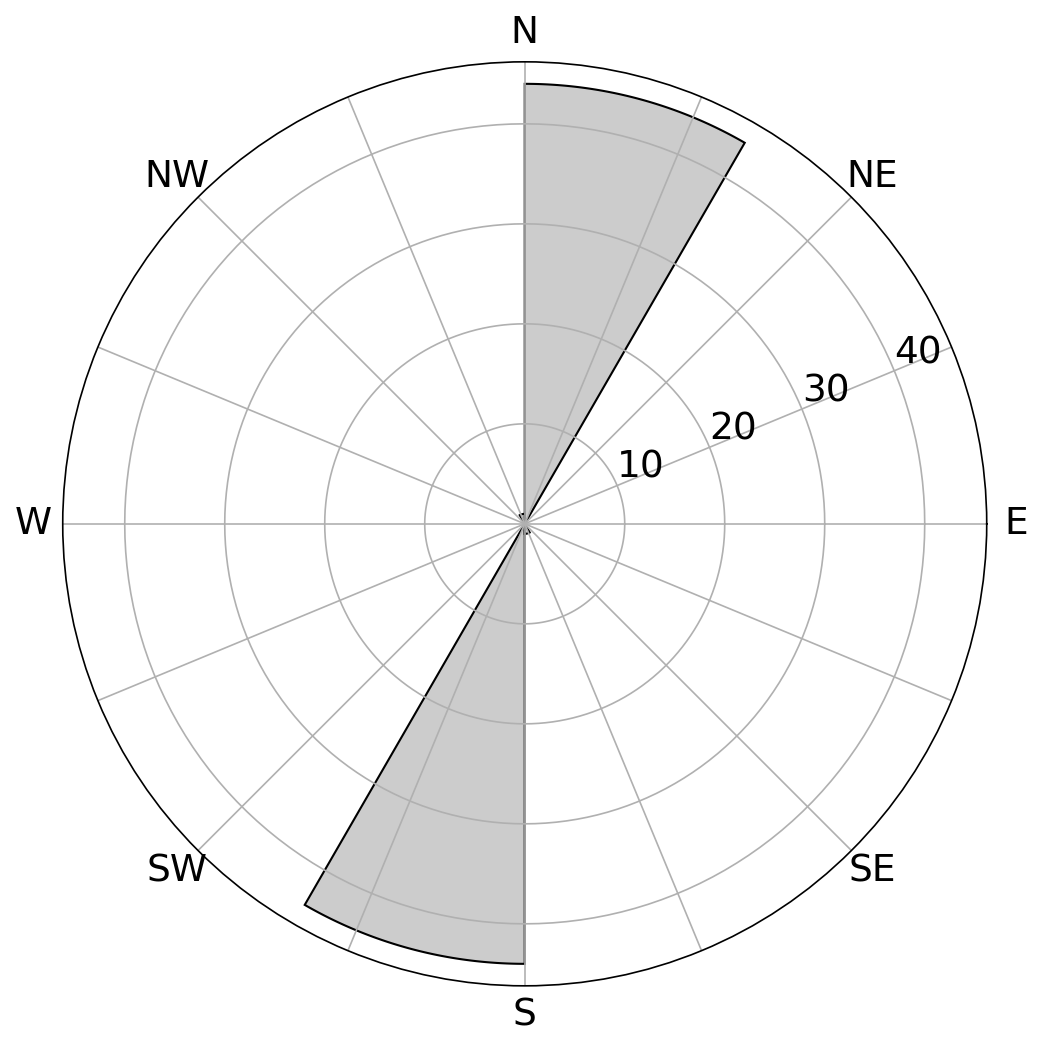}
    \caption{north (top)}\label{CrackTop}
\end{subfigure}
\caption{Windrose diagrams for different crack orientations and cracks attached to \textbf{a)} the W-face, \textbf{b)} the E-face and to \textbf{c)} the north.}\label{SameCrackLengthLeftrightTop}
\end{figure}

Figure \ref{CrackLeft} shows the windrose diagram for the case of a crack attached to the W-face. Computed crack propagation directions are preferentially oriented in the North-West to South-East (NW-SE) direction, regardless of the initial different orientations of the cracks taken into account. Furthermore, a minimum number of computed crack propagation directions are oriented East (E) to West (W). 
In Fig. \ref{Crackright} it is shown the windrose diagram for the case of a crack attached to the E-face. In this case the preferential orientation of the computed crack propagation directions is in the North-East to South-West (NE-SW) direction. There are also few cracks propagating in a direction aligned  East (E) - West (W). 
The last case, where the crack is attached to the north (top) is shown in Fig. \ref{CrackTop}. According to the windrose diagram, the preferential orientation of the computed crack propagation directions is in the North to South (N-S) direction.

\begin{figure}[h!]
\centering
\includegraphics[width=0.57 \textwidth]{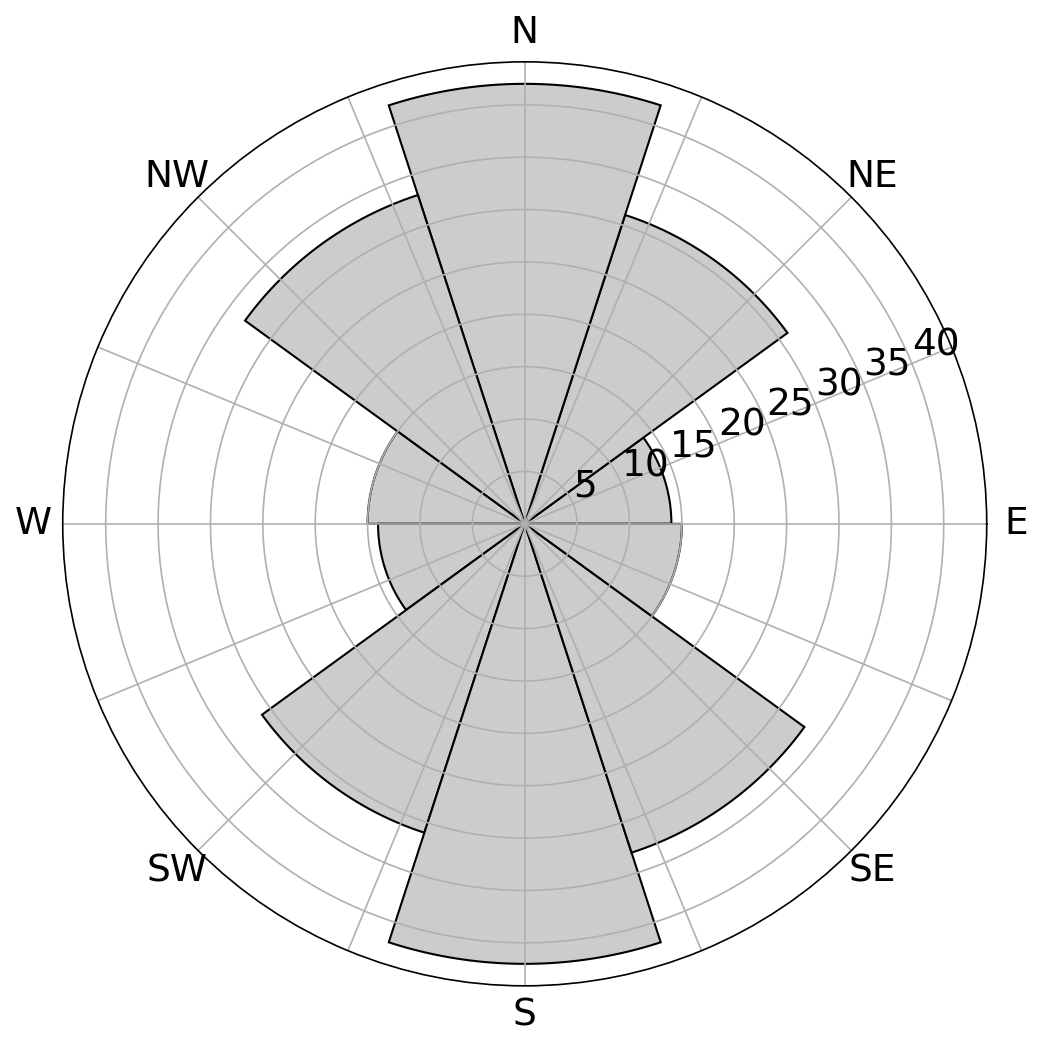}
\caption{Windrose diagram gathering the cases where crack is attached to the W-face, to the E-face and to the north (top) for different crack orientations.}
\label{CrackTipDifferentPosition}
\end{figure}

Finally in Fig. \ref{CrackTipDifferentPosition} it is presented a windrose diagram that gathers all the cases described above. In summary, the distribution of all the computed crack propagation directions are preferentially oriented in a higher concentration in the North to South (N-S), in the North-West to South-East (NW-SE) and in the North-East to South-West (NE-SW) directions as it was already described.

It is noted that the cracks simulated up to now have a length considerably smaller than the ones observed on the boulders on the surface of Bennu \citep{Molaro2020NatCo..11.2913M,Lauretta2019Natur.568...55L,DellaGiustina2019NatAs...3..341D,Walsh2019NatGe..12..242W,Delbo2019EPSC...13..176D}. 
In the following, the crack propagation direction is studied as a function of the initial crack size. From the different configurations shown in Fig. \ref{GeometryBeamAsteroidBennu}, it was selected the case for the crack attached to the E-face. For this, crack propagation direction is computed by placing the crack tip at 3 different positions inside the domain of the a-face; for each one of this positions 3 different lengths of the a-face were used; for each one of these cases the orientation of the crack was varied as it was done in the previous simulations.

we varied the orientation of the crack as it was done in the previous simulations.

\begin{figure}[h!]
\centering
\begin{subfigure}[b]{0.32\textwidth}
    \centering
    \includegraphics[width=\textwidth]{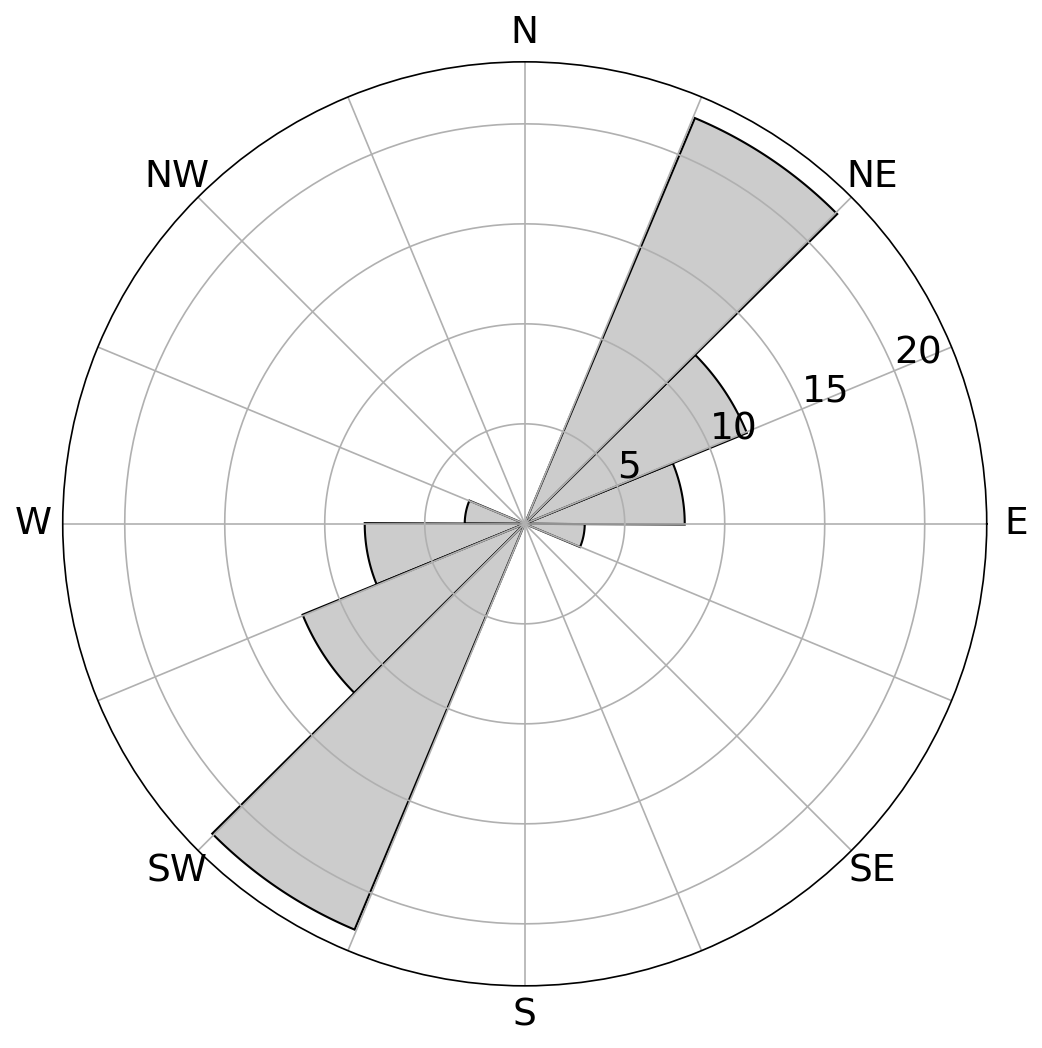}
    \caption{a-face length: 200 mm}\label{Crack_Length_2_mm_Length_1}
\end{subfigure}
\begin{subfigure}[b]{0.32\textwidth}
    \centering
    \includegraphics[width=\textwidth]{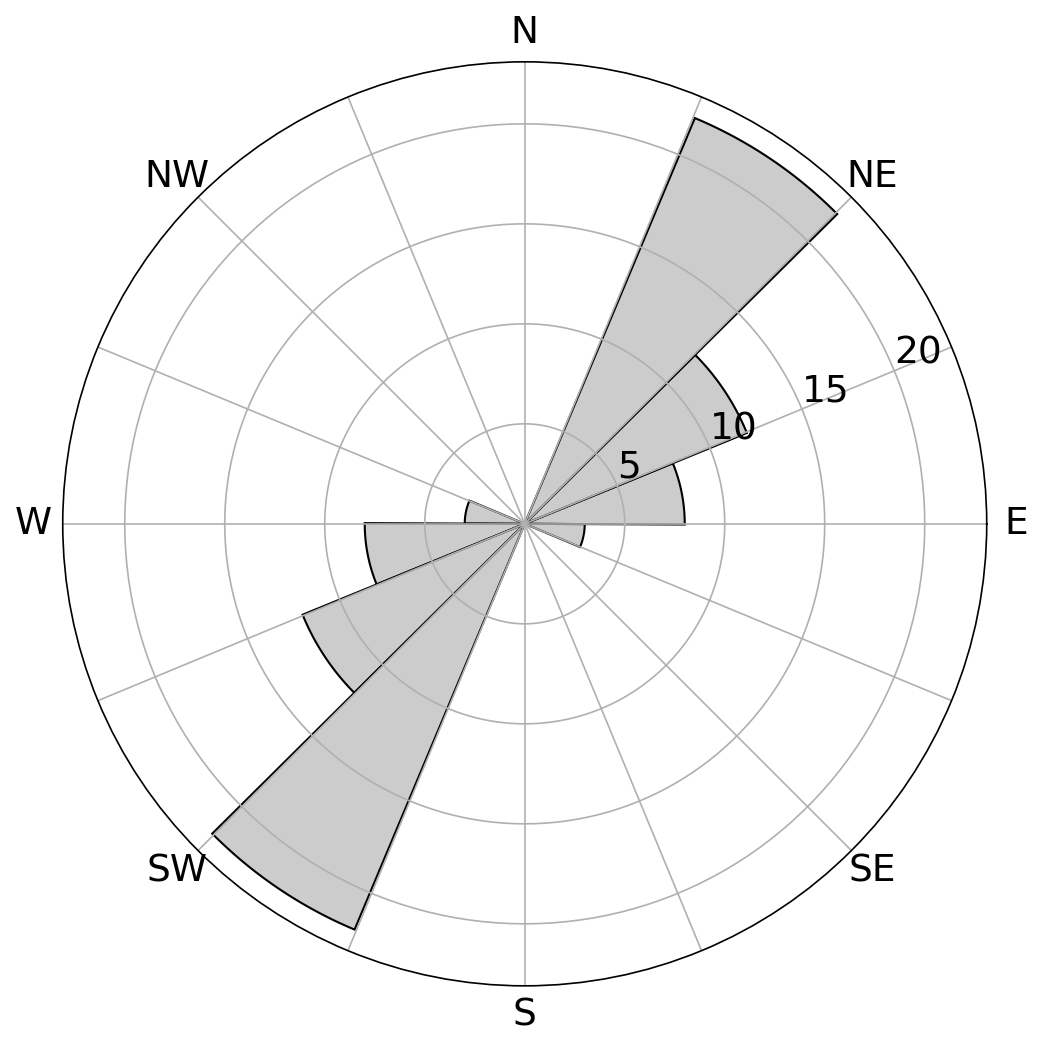}
    \caption{a-face length: 500 mm}\label{Crack_Length_2_mm_Length_2}
\end{subfigure}
\begin{subfigure}[b]{0.32\textwidth}
    \centering
    \includegraphics[width=\textwidth]{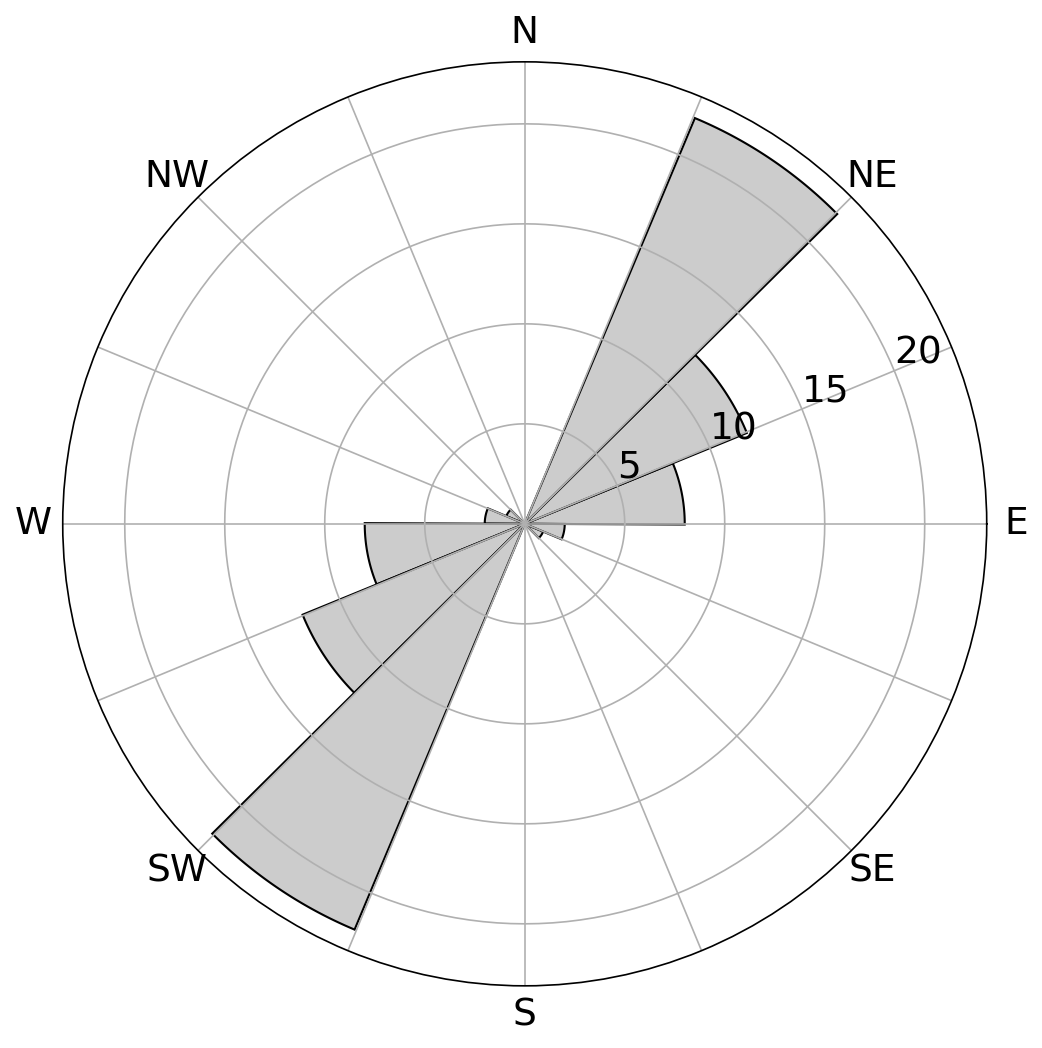}
    \caption{a-face length: 1000 mm}\label{Crack_Length_2_mm_Length_3}
\end{subfigure}
\caption{Windrose diagrams for a crack attached to the E-face when crack length equals to $\frac{2}{\mid\sin azimuth\mid}$ $mm$ and the length of the beam was varied for different crack orientations.}\label{SameCrackLength1VaryingBeamLength}
\end{figure}

Figure \ref{SameCrackLength1VaryingBeamLength} shows that for a small crack length ($\frac{2}{\mid\sin azimuth\mid}$ $mm$), increasing the length of the a-face does not play an important role in the distribution of the computed crack propagation directions. Computed crack propagation directions are preferentially oriented in the North-East to South-West (NE-SW) direction, with a few amount of computed crack propagation directions going to the East (E) and to the West (W).

\begin{figure}[h!]
\centering
\begin{subfigure}[b]{0.32\textwidth}
    \centering
    \includegraphics[width=\textwidth]{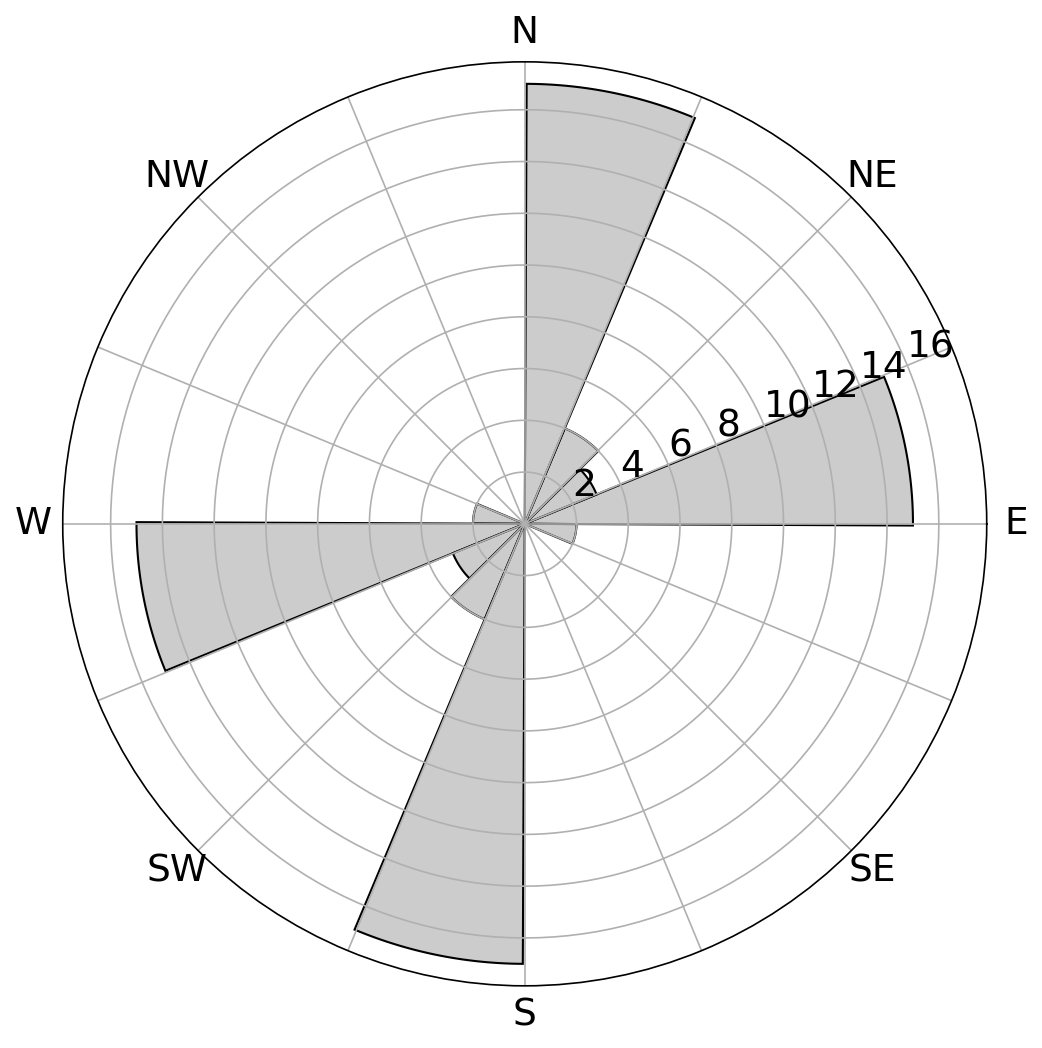}
    \caption{a-face length: 200 mm}\label{Crack_Length_20_mm_Length_1}
\end{subfigure}
\begin{subfigure}[b]{0.32\textwidth}
    \centering
    \includegraphics[width=\textwidth]{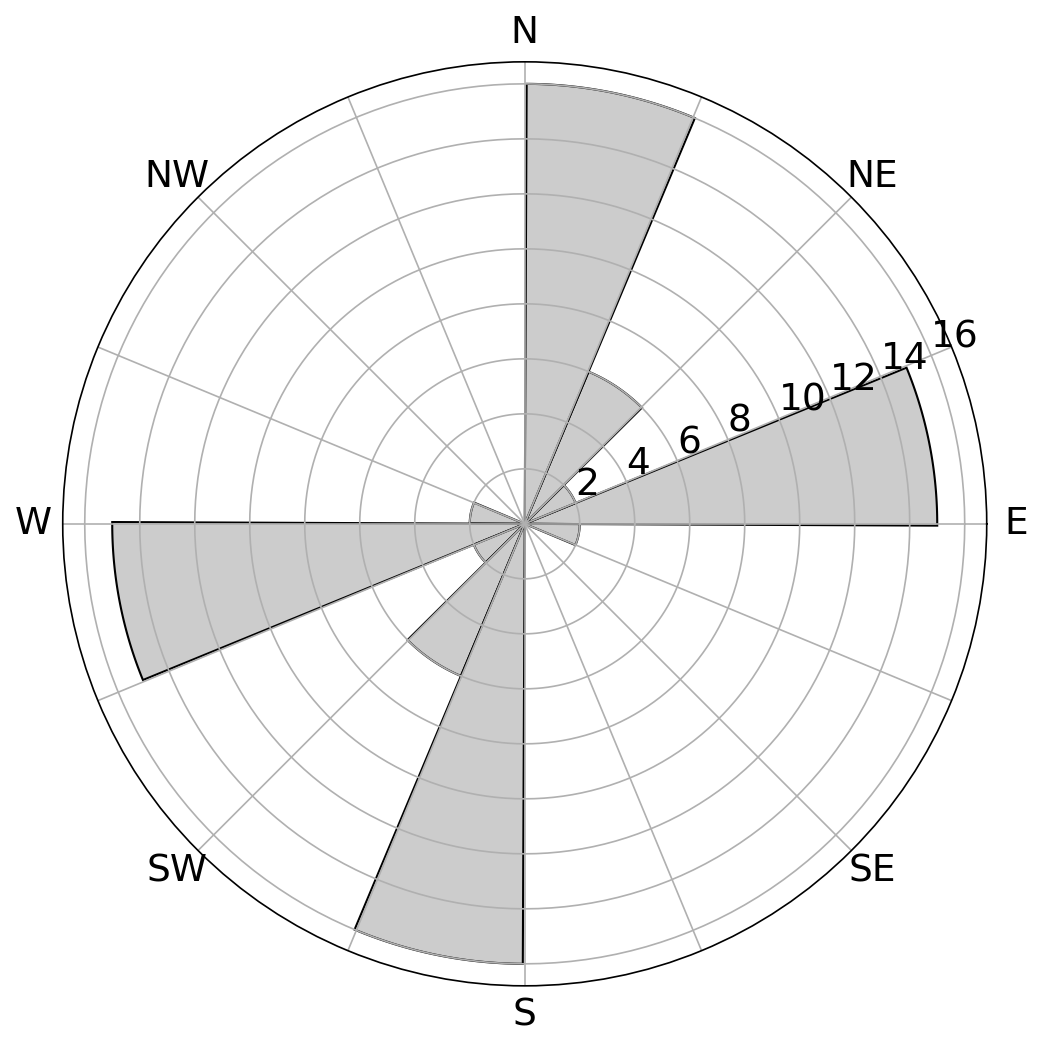}
    \caption{a-face length: 500 mm}\label{Crack_Length_20_mm_Length_2}
\end{subfigure}
\begin{subfigure}[b]{0.32\textwidth}
    \centering
    \includegraphics[width=\textwidth]{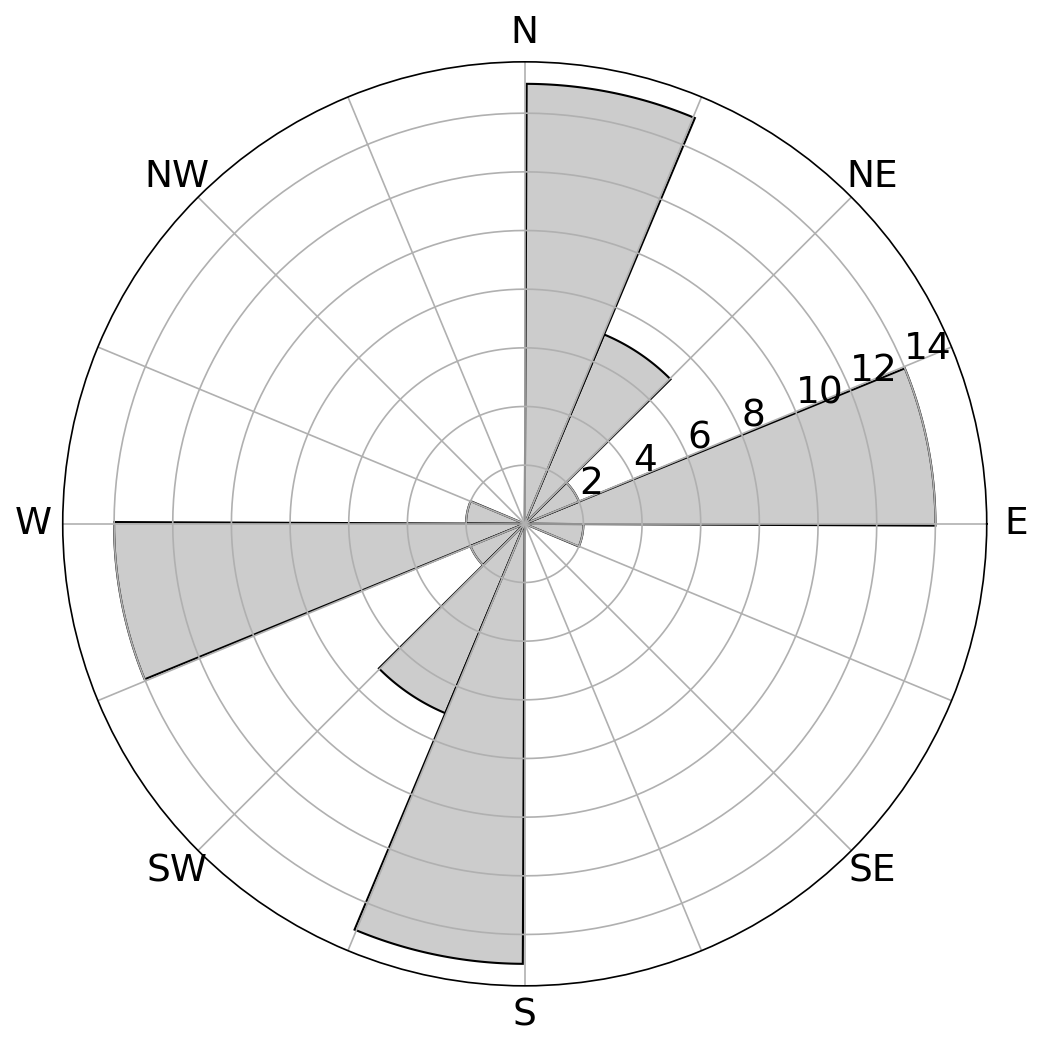}
    \caption{a-face length: 1000 mm}\label{Crack_Length_20_mm_Length_3}
\end{subfigure}
\caption{Windrose diagrams for a crack attached to the E-face when crack length equals to $\frac{20}{\mid\sin azimuth\mid}$ $mm$ and the length of the beam was varied for different crack orientations.}\label{SameCrackLength2VaryingBeamLength}
\end{figure}

In Fig. \ref{SameCrackLength2VaryingBeamLength}, it is possible to see that when increasing the length of the a-face for cracks whose lengths are $\frac{20}{\mid\sin azimuth\mid}$ $mm$, the distribution of the computed crack propagation directions oriented in the North to South (N-S) direction decreases, while the cracks oriented in the North-East to South-West (NE-SW) direction increase. It should also be noted that the amount of cracks directed to the East (E) and the West (W) decreases when increasing the length of the a-face from 500 $mm$ to 1000 $mm$.

\begin{figure}[h!]
\centering
\begin{subfigure}[b]{0.46\textwidth}
    \centering
    \includegraphics[width=\textwidth]{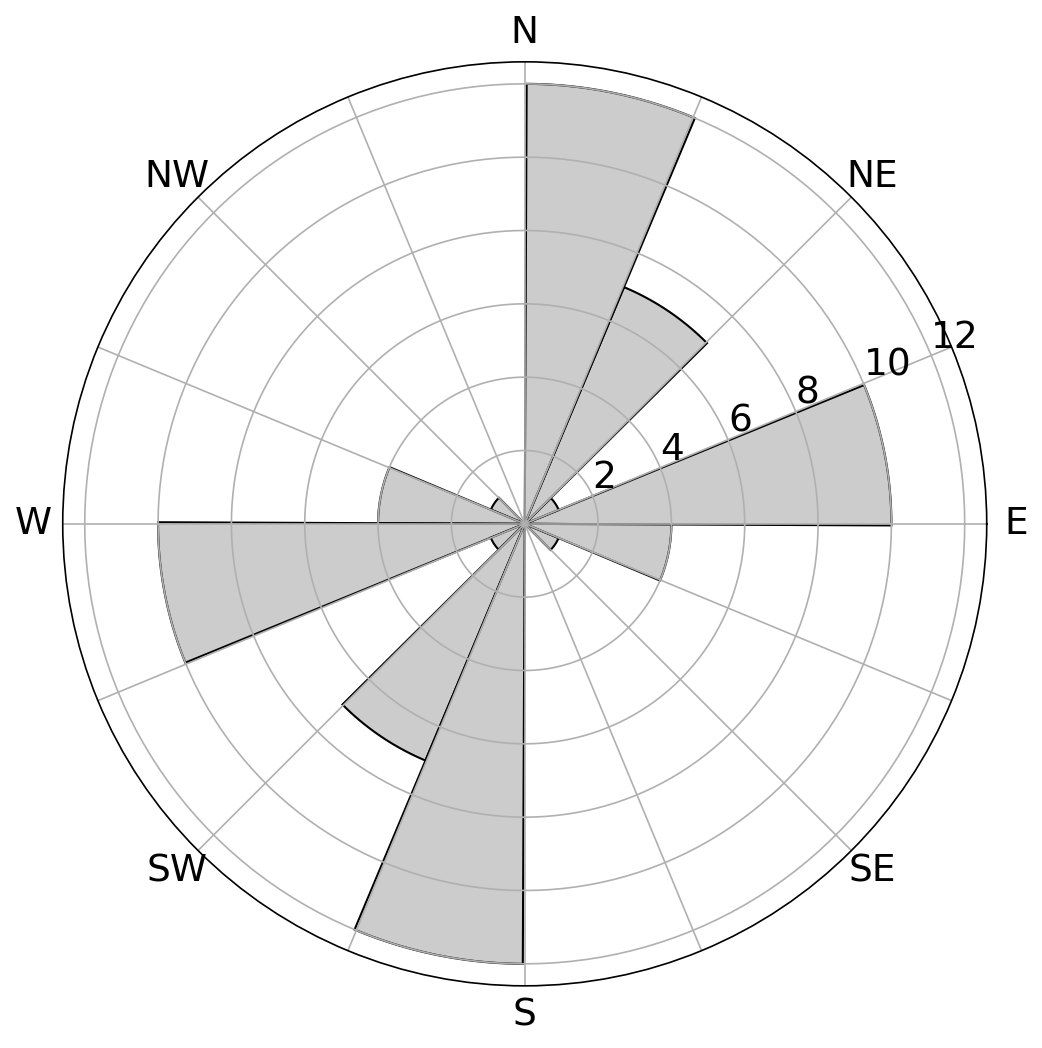}
    \caption{a-face length: 500 mm}\label{Crack_Length_200_mm_Length_1}
\end{subfigure}
\begin{subfigure}[b]{0.46\textwidth}
    \centering
    \includegraphics[width=\textwidth]{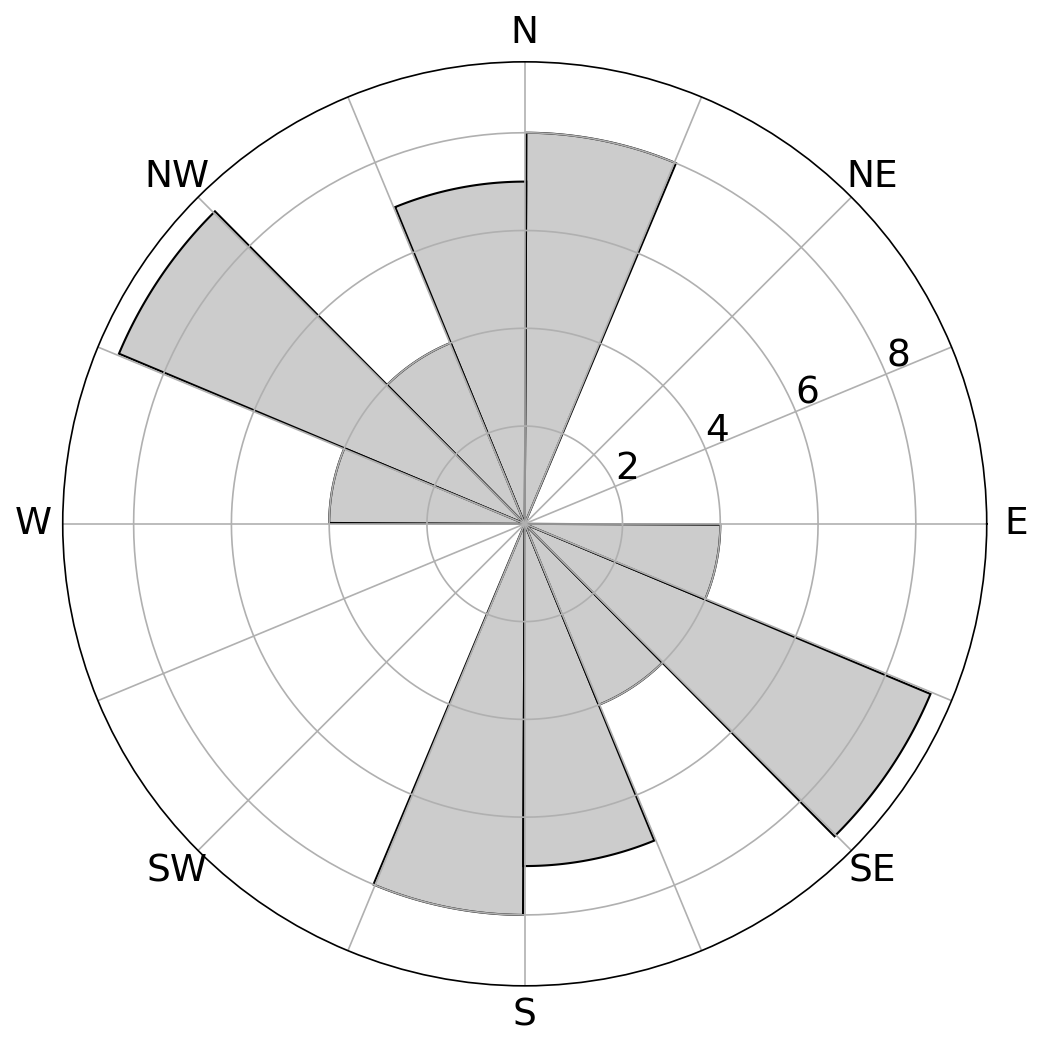}
    \caption{a-face length: 1000 mm}\label{Crack_Length_200_mm_Length_2}
\end{subfigure}
\caption{Windrose diagrams for a crack attached to the E-face when crack length equals to $\frac{200}{\mid\sin azimuth\mid}$ $mm$ and the length of the beam was varied for different crack orientations.}\label{SameCrackLength3VaryingBeamLength}
\end{figure}

Finally, in Fig. \ref{SameCrackLength3VaryingBeamLength} it is presented the windrose diagrams for a configuration where crack length is equal to $\frac{200}{\mid\sin azimuth\mid}$ $mm$. Taking into account that the length of the a-face took the values of 200 $mm$, 500 $mm$ and 1000 $mm$, for this last case, the configuration of the a-face with a length equal to 200 $mm$ was not simulated. Fig. \ref{SameCrackLength3VaryingBeamLength} shows that when increasing the length of the a-face from 500 $mm$ to 1000 $mm$, a redistribution of the computed crack propagation directions takes place. The initially cracks preferentially oriented in the North-East to South-West (NE-SW) direction changed their orientation into the North-West to South-East (NW-SE) direction.

Figure \ref{StressFieldCrackTip} shows the stress field through the maximum principal stress for three cracks with different orientations: attached to the W-face and directed to the East, attached to the north face of the boulder and directed to the South, and attached to the E-face of the boulder and directed to the West. It is worth noting that in all the cases, the computed crack propagation direction ({dashed black line), the one that maximises $G$, is perpendicular to the direction in which the maximum principal stress takes place (dashed white line). For all cases, the maximum stress near the crack tip over a temperature cycle exceeds some MPa, which is a significant fraction of the typical strength of carbonaceous chondrites and even comparable to the strength of boulders on Bennu \citep{Ballouz2020nat}. This indicates that cracks propagate under these stresses. Although it should be noted that, theoretically, the stress at the crack tip is supposed to be almost infinity ($\infty$), it depends a lot on the mesh size. The use of quarter-point elements \citep{Barsoum1976} is recommended for better accuracy of the stress field.

\begin{figure}[h!]
\centering
\includegraphics[width=0.78 \textwidth]{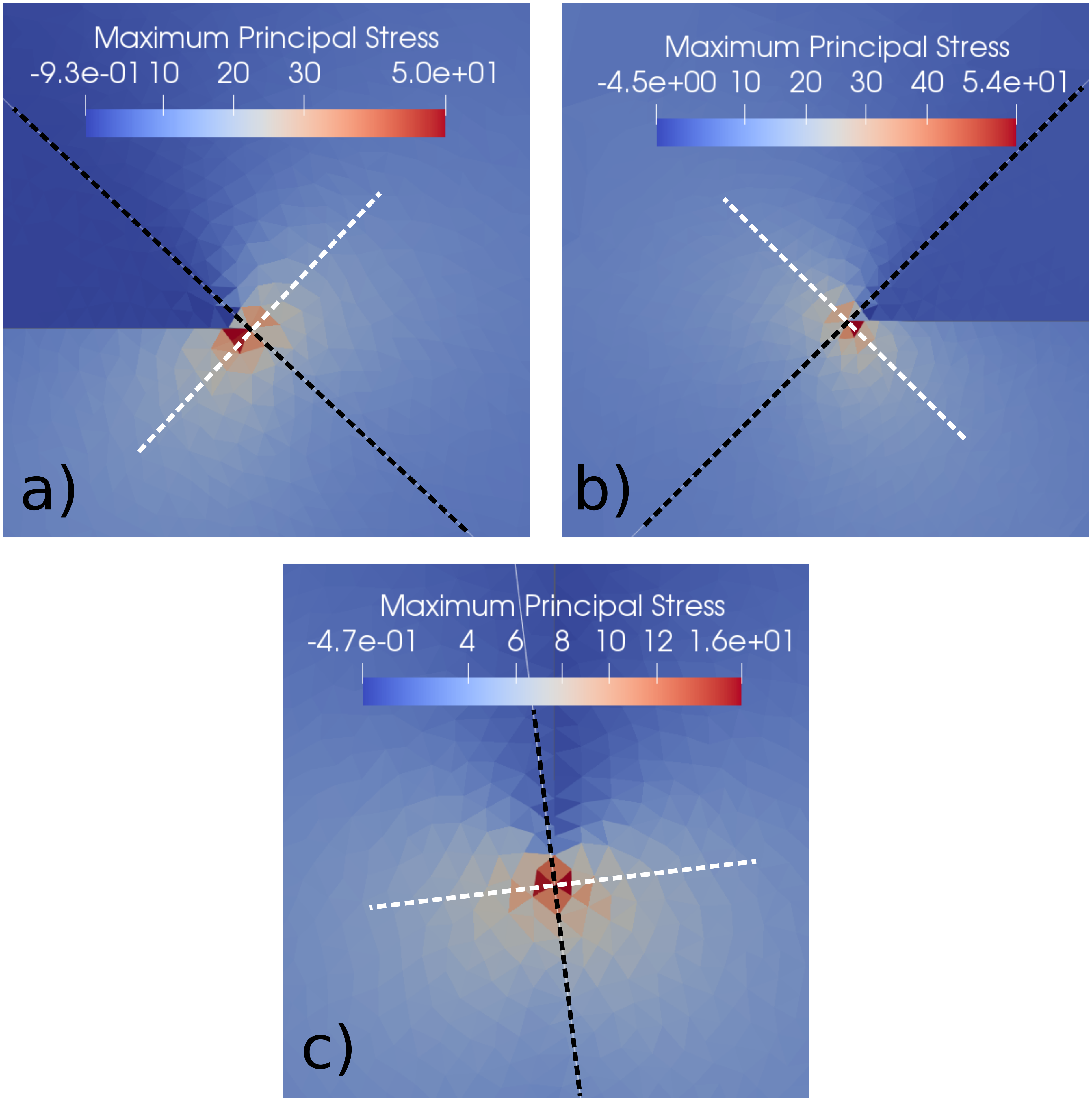}
\caption{Maximum principal stress $(MPa)$ for a crack attached to \textbf{a)} the W-face, \textbf{b)} the E-face and to \textbf{c)} the north (top). In all cases, the dashed black line represents the computed crack propagation direction, while the dashed white line represents the axis associated with the maximum principal stress.} 
\label{StressFieldCrackTip}
\end{figure}

To study whether the crack is propagating it can also be used the typical approach of thermal fatigue and determine the crack growth length per cycle using the Paris' law. To do so, the configuration in which the crack is attached to the W-face was chosen. Additionally, it was selected an orientation of the crack (azimuthal angle) equal to $90^{\circ}$, i.e. pointing to the east. In this case, as the maximum energy release rate ($G_{max}$) is known, and assuming plane strain conditions, it is possible to compute the maximum stress intensity factor ($K_{Imax}$) using equation (\ref{EquivalenceKG}). In this case, $G_{max}$ in the thermal cycle is equal to $5.15 \ \times \ 10^{-4} \ [MPa \ . \ mm]$, therefore, using equation (\ref{EquivalenceKG}), $K_{I}$ is equal to $4.95 \ [MPa \ . \ mm^{0.5}]$. Due to the fact that in this work it is dealt with stresses generated by thermal cycles, the crack tip is subjected to both tensile and compressive stresses over a full cycle. Thus, ignoring crack closure \citep{suresh_1998}, the lowest stress intensity factor experienced by the crack tip  is simply zero. According to this, it follows that $\Delta K_{I}$ is equal to $4.95 \ [MPa \ . \ mm^{0.5}]$. For the case where constant crack growth is assumed, it can be computed that for the presented thermal cycle, the crack growth rate is $2.42 \ \times \ 10^{-4} \ [mm \ . \ cycle^{-1}]$, which is of the order of $0.5 \ [mm \ . \ yr^{-1}]$ (or $\sim0.5 [m]$ in thousand years), in good agreement with previous studies \citep{Delbo2014Natur.508..233D}. This indicates that cracks would propagate on Bennu's boulders solely due to diurnal  thermal stresses. It is cautioned here that crack propagation is a non linear process, where the rate of propagation is among other parameters a function of the position of the crack tip with respect to the temperature gradient. This means that a full simulation of the crack growth from the beginning until its size is comparable to the size of the hosting boulder should be carried out in order to estimate the time required to fracture the boulder, which is beyond the scope of this paper.

\section{Discussion}
\label{DiscussionSection}

We begin to notice that in Fig. \ref{CrackLeft}, \ref{Crackright} and \ref{CrackTop}, the range of computed crack propagation directions is preferentially concentrated in some sectors of the windrose diagrams. This could be explained by the fact that when a crack orientation is defined, the range of possible crack propagation directions, is determined by the limit angle corresponding to pure shear defined in the work of \cite{erdogan_crack_1963}. This value is approximately $\pm 70^{\circ}$ with respect to the crack axis orientation. It should also be noted that, the crack was only placed at three different places inside the domain, i.e. attached to the East, the West faces and attached to the North rim of the domain. This means that not all the angles belonging to the range $[0^{\circ}, 360^{\circ}]$ were taken into account. For example, in Fig. \ref{Crackright}, i.e., for the case of a crack attached to the E-face, the computed crack propagation directions range from $202.5^{\circ}$ to $292.5^{\circ}$. From this figure, it is worth noting that most of the computed direction are oriented to the South-West (SW), and due to the already explained symmetry of the wind rose, the preferential orientation of the computed crack propagation directions is in the North-East to South-West (NE-SW) direction.

When plotting  the windrose diagrams for the three evaluated cases (crack attached to the E-, W-face and the north rim of the domain) in the same graph, as shown in Fig. \ref{CrackTipDifferentPosition}, the sector grouping the majority of the computed directions is the one oriented in the North to South (N-S) direction, followed by the groups oriented in the North-West to South-East (NW-SE) and in the North-East to South-West (NE-SW) directions, respectively. Looking at the results of the three simulated cases, it is worth noting that cracks on asteroid Bennu propagate mainly in the North-South (N-S) direction.

In Fig. \ref{StressFieldCrackTip} it should be noticed that the computed crack propagation directions in the three cases presented there, are perpendicular to the axis of the maximum principal stresses, as stated by the literature. This makes us feel confident about the results coming out from the coupled thermoelastic model with the linear elastic fracture mechanics approach that was implemented in this work. Additionally, it is worth mentioning that our model to compute crack propagation direction was validated in previous work \citep{Uribe_2020}.

From Fig. \ref{SameCrackLength1VaryingBeamLength} it can be noticed that for short crack lengths, the increasing of the length of the a-face does not highly affect the tendency of the distribution of the computed crack propagation directions. Cracks mainly continue propagating oriented in the North-East to South-West (NE-SW) direction. When increasing the length of the a-face from 500 $mm$ to 1000 $mm$, a few amount of cracks changed its orientation into the North-West to South-East (NW-SE) direction. It tells us that even if the temperature gradient along the a-face changes, cracks of short length are not affected.

According to Fig. \ref{SameCrackLength2VaryingBeamLength}, when crack length increases, a redistribution of the computed directions takes place. It is possible to see two main groups, one oriented almost in the North to South (N-S) direction and another one oriented almost in the West to East (W-E) direction. Meanwhile, for the case when the crack length is very long compared with the length of the a-face, as depicted in Fig. \ref{SameCrackLength3VaryingBeamLength}, a marked tendency is clearly observed. Computed crack propagation directions are preferentially oriented in the North-West to South-East (NW-SE) direction, which is consistent with the observations performed by \citet{Delbo2019EPSC...13..176D}.

Taking into account that the duration of the thermal cycle used in this work is 4.3 $h$, and assuming a constant crack propagation rate, the crack growth rate computed here, whose value is $2.42 \ \times \ 10^{-4} \ [mm \ . \ cycle^{-1}]$, is equivalent to a crack growth rate of about $0.5 \ [mm \ . \ year^{-1}]$. This value is not quite different from the one measured by \citet{Delbo2014Natur.508..233D} through laboratory experiments on two meteorites: a carbonaceous chondrite and an ordinary chondrite. However, as stated in \citet{Delbo2014Natur.508..233D}, the crack propagation rate is typically a nonlinear function between the crack size, the maximum variation of the stress intensity factor over the whole cycle and material properties. However, considering that this work is a first attempt to describe crack growth due to thermal fatigue, the results are promising.

\section{Conclusion}
\label{ConclusionSection}

An approach to compute 2D crack propagation direction under the presence of thermal gradients has been presented. It combines thermoelasticity and linear elastic fracture mechanics theories in order to compute crack propagation directions driven by diurnal temperature cycling and under conditions similar to those existing on the asteroid Bennu. Through the different scenarios simulated in this work, the robustness and the accuracy of the proposed approach in terms of crack propagation direction is shown.

It is found that the distribution of the computed crack propagation directions for different configurations on a simple domain simulating Bennu's boulders, show a preferential direction from North to South (N-S) and from North-East to South-West (NE-SW) for shorter cracks. For longer cracks, the preferential direction is from North to South (N-S) and from North-West to South-East (NW-SE). This conclusion is supported by observations performed by \cite{Delbo2019EPSC...13..176D}.

According to the results obtained by means of our thermomechanical model combined with a well-known fatigue model (Paris's law) and assuming constant crack growth length, it is found that on asteroid Bennu cracks grow at rate approximately equal to $0.5 \ [mm \ . \ year^{-1}]$. This is a value very similar to the one found on experiments perform in the work of \citet{Delbo2014Natur.508..233D}, giving us a good insight of the promising approach proposed. 

However, it is cautioned that crack propagation is a highly non-linear process. In order to estimate the time to fracture Bennu's meter-sized rocks a full-blown simulation shall be performed to calculate the crack growth rate at different crack growth stages. This is really important because crack growth rate highly depends on the geometry and on the stress intensity factor at the crack tip, and those will change as the crack tip evolves trough the domain.

\section*{Acknowledgements}
We acknowledge support from Academies of Excellence on Complex Systems and Space, Environment, Risk and Resilience of the Initiative d'EXcellence "Joint, Excellent, and Dynamic Initiative" (IDEX JEDI) of the Universit\'e C\^ote d'Azur and the  Center for Planetary Origin (C4PO) A Material approach (\url{https://www-n.oca.eu/morby/C4PO/C4PO-home.html}). We also acknowledge support from The DIGIMU ANR industrial chair (\url{https://chaire-digimu.cemef.mines-paristech.fr/}) as well as support from the French space agency CNES.

\addcontentsline{toc}{section}{Bibliography}

\bibliographystyle{unsrtnat}

\bibliography{FinalBiblio}

\end{document}